\documentstyle[11pt,aaspp4]{article}

\def\etal{{\it et~al.\ }}
\def\eg{{e.g.,}\ }
\def\ie{{i.e.,}\ }
\def\etal{{et al.}\ }
\def\parcsec{{\tt ''}\mskip -7.6mu.\,}

\def\kms{{km~s$^{-1}$}}
\def\simgt{\ {\raise-.5ex\hbox{$\buildrel>\over\sim$}}\ }
\def\simlt{\ {\raise-.5ex\hbox{$\buildrel<\over\sim$}}\ }

\begin{document}

\title{Chemical Abundances of Planetary Nebulae in the Bulge and Disk of M31}

\author{George H. Jacoby\altaffilmark{1}}
\affil{Kitt Peak National Observatory, National Optical Astronomy Observatories, \\
 P.O. Box 26732, Tucson, AZ, 85726}
\authoremail{gjacoby@noao.edu}

\altaffiltext{1}{Visiting Astronomer, Steward Observatory, 
     University of Arizona, 933 N. Cherry Ave, Tucson, AZ 85721}

\author {Robin Ciardullo\altaffilmark{2}}
\affil{Department of Astronomy and Astrophysics, Penn State University,
525 Davey Lab, University Park, PA 16802}
\authoremail{rbc@astro.psu.edu}

\altaffiltext{2}{Visiting Astronomer, Kitt Peak National Observatory, 
     National Optical Astronomy Observatories, which is operated by the
     Association of Universities for Research in Astronomy, Inc., under
     cooperative agreement with the National Science Foundation.}

\begin{abstract}
We derive abundances and central star parameters for 15 planetary
nebulae (PNe) in M31: 12 in the bulge and 3 in a disk field 14 kpc from
the nucleus.  No single abundance value characterizes the bulge stars:
although the median abundances of the sample are similar to those seen for
PNe in the LMC, the distribution of abundances is several times broader,
spanning over 1 decade.  None of the PNe in our sample approach the super
metal-rich ([Fe/H] $\sim +0.25$) expectations for the bulge of M31,
although a few PNe in the sample of Stasi\'nska, Richer, \& Mc~Call
(1998) come close.  This [O/H] vs [Fe/H] discrepancy is likely due to
a combination of factors, including an inability of metal-rich stars
to produce bright PNe, a luminosity selection effect, and an abundance
gradient in the bulge of M31. We show that PNe that are near the bright
limit of the [O~III] $\lambda 5007$ planetary nebula luminosity function
(PNLF) span nearly a decade in oxygen abundance, and thus, support
the use of the PNLF for deriving distances to galaxies (Jacoby 1996)
with differing
metallicities. We also identify a correlation between central star mass
and PN dust formation that partially alleviates any dependence of the
PNLF maximum magnitude on population age.  Additionally, we identify a
spatially compact group of 5 PNe having unusually high O/H; this subgroup
may arise from a recent merger, but velocity information is needed to
assess the true nature of the objects.
\end{abstract}


\section{Introduction}
Ever since the pioneering work by Spinrad \& Taylor (1971) on the optical
spectra of M31's bulge, the evidence from integrated spectroscopy has
been that the cores of giant E/S0 galaxies and the bulges of large
spirals contain a significant population of metal-enriched stars (e.g.,
Bica, Alloin, \& Schmidt 1990; Worthey, Faber, \& Gonzalez 1992 (WFG);
Davies, Sadler, \& Peletier 1993; Carollo, Danziger, \& Buson 1993;
Vazdekis \etal 1997).  For example, WFG reaffirm the Spinrad \& Taylor
(1971) result that the stars in the bulge of M31 have [Fe/H] $\sim +0.23$,
i.e., a value 70\% higher than the Sun.

On the other hand, Ferguson \etal (1991) have shown that the very weak
lines in the UV upturn of NGC 1399 are consistent with the existence of a
metal-poor population mixed-in with the metal-rich stars.  This idea is
confirmed is M32, where the color-magnitude diagram of Freedman (1989)
demonstrates a large range ($\sim 0.6$ dex) in the metallicity of the stars,
and the HST results by Grillmair \etal (1996) confirm an
even wider range ($\sim 1.1$ dex).
Furthermore, studies of the stars in the bulge of our own Galaxy have
shown that there can be large abundance variations from star to star,
and that extreme care must be taken when interpretating even these
relatively simple data (see review by McWilliam 1997).  Given this
complication, our ability to investigate the chemistry of galaxies via
integrated spectroscopy seems limited.  Instead, studies of large numbers
of individual stars are preferred.

Planetary nebulae (PNe) are excellent tools for obtaining both the mean
metallicity and metallicity dispersion within large, nearby galaxies.
Because bright PNe are plentiful, large numbers of objects are available
for analysis.  Moreover, although internal chemical processing does
affect some of a PN's constituent elements (e.g., helium, nitrogen, and
carbon), important elements, such as oxygen, neon, argon, and sulfur are
generally considered to be largely unaffected.  Of greatest importance, however,
is the presence of planetary nebulae throughout a galaxy.  Because PNe do
not belong to any single stellar population, they can be used to measure
bulge, disk, and halo abundances via a single analysis technique.
Abundance gradients can therefore be measured from the very center of
the galaxy to the outermost regions in a uniform manner (cf.~Jacoby
\& Ford 1986), rather than the mixture of techniques (absorption line
measurements for bulge stars, emission line measurements for disk H~II
regions, broadband colors for halo stars) that are otherwise necessary.

The power of planetary nebulae to probe both the make-up and chemical
history of a stellar population has been demonstrated by Dopita \etal (1997).
By deriving the central star properties ($T_{eff}$, L/L$_{\odot}$, and
M/M$_{\odot}$), and adopting an initial-to-final mass relation (e.g.,
Vassiliadis \& Wood 1994), Dopita \etal were able to compare the alpha
element abundances of LMC planetaries to the main sequence turn-off
ages of their progenitors, and consequently, to trace the
history of chemical enrichment within that galaxy.  While mergers and infall 
events may confuse the issue of chemical enrichment in larger, more complex 
systems, the degree to which a monotonic relation between abundance and
age does exist can place limits on the importance of mergers in a galaxy's
evolution.

Nevertheless, aside from observations of objects in the Magellanic Clouds,
there is little spectroscopic data for extragalactic planetary nebulae.
In particular, there have been only two previous studies of planetary
nebulae in M31: the early efforts of Jacoby \& Ford (1986), and the recent
observations of Richer, Stasi\'nska, \& Mc~Call (1998; hereafter RSM).
The former study included only three PNe, two in the halo of M31, and
one in the disk; the latter work consisted of observations of 28 bulge
PNe and 2 outer disk PNe.  Based on their observations, Stasi\'nska, 
Richer, \& Mc~Call (1998; hereafter SRM) find that
a very wide range of abundances exists among the M31 bulge stars, with
[O/H] ranging between $-1.0$ and $+0.2$.

%As a by-product of the abundance analysis of PNe in other galaxies,
%one derives the extinction to the nebula. This parameter, in turn,
%can be used to correct the nebula's location in the planetary nebula
%luminosity function (PNLF). The PNLF has been used extensively to derive
%distances to galaxies (Jacoby 1996), and one of the larger contributors
%to the uncertainty in those distances is the definition of the reference
%PNLF (Ciardullo \etal 1989).  In a similar manner, the PNLF method can
%be examined as a function of chemical composition once many PNe in a
%mixed population at a fixed distance are observed.  Initial studies
%suggest a very weak dependence on metallicity (Ciardullo \& Jacoby 1992;
%Richer, McCall, \& Arimoto 1997; Kennicutt \etal 1998).

In this paper we report spectroscopic observations and model nebula
analyses for 15 PNe in M31.  In \S 2, we describe our observations and
reductions; in \S 3, we use photoionzation models to derive abundances
and central star parameters for our objects.  We review the abundance
results for PNe in M31's bulge and disk in \S 4, and we discuss the PN
abundances in the context of their metal-rich environment in \S 5.

\section{Observations and Reductions}

\subsection{Observing strategy}

We observed 16 planetary nebulae in the bulge and disk of M31 with
slit masks on the R-C spectrograph of the Kitt Peak 4-m telescope.
The spectrograph configuration consisted of a 316 lines/mm grating
blazed at 4000 \AA\ in first order (KPC-10A), a WG360 blocking filter,
and the T2KB $2048 \times 2048$ CCD with 24~micron pixels.  This setup,
combined with an atmospheric dispersion corrector and $2\arcsec$ wide
slitlets (multiple, short slits, etched into a focal plane mask), 
enabled us to perform relative spectrophotometry over a full
octave of wavelength from 3700~\AA\ to 7400~\AA, with 6.9~\AA\ resolution
at 2.75~\AA\ per pixel.  Perpendicular to dispersion, the instrument
delivered a scale of $0\parcsec 69$ per pixel, over a slit length of
$\sim 12\arcsec$. With complete optical coverage over a very large range of
airmasses, this setup is very good for measuring nebular abundances;
it is, however, non-optimal for deriving radial velocities because of
the low dispersion and wide slit.

Observations were taken on two nights.  On UT date 1995 September 20,
we observed 13 of the bulge PNe identified by Ford \& Jacoby (1978) and
Ciardullo \etal (1989; CJFN) in their surveys of M31's bulge and inner
disk.  These observations consisted of 13 1800-second exposures, taken
under photometric conditions with a typical seeing $\sim 0\parcsec 9$.
The following night, we observed 3 PNe found by Ford \etal (2000; FJCHP) in
their large area survey of M31's disk.  These objects, which are at a
projected galactocentric distance of 14~kpc, were observed via 12 1800-sec
exposures taken through light cirrus and $\sim 0 \parcsec 9$ seeing.
The data of both nights were calibrated via a series of exposures of
the spectrophotometric standards BD+26 3941, G191B2B, HD~192291, and
BD+28~4211 (Stone 1977; Oke 1974), with the stars positioned in each
slitlet at least once.  A summary of the observations appears in Table~1.
Note that the PNe chosen for study span a range of 2.4~mag in the [O~III]
$\lambda 5007$ planetary nebula luminosity function.  Kaler (1995)
noted that the spectra of the brightest M31 bulge PNe exhibit considerable
uniformity, and so, by including
fainter objects in our sample, we sought to decrease the impact of
selection effects on the derived distribution of nebular abundances.

\subsection{Special data reduction issues}

The need for both wide wavelength coverage and accurate sky subtraction of the 
bright galaxy background presented a unique set of problems for the reduction
of our slitlet data.  The design of the R-C spectrograph is such that the
instrument's focus changes at the ends of the spectral range.  
Consequently, the width of our spectra varied with wavelength,
especially for those objects positioned near the edge of the instrument's
field-of-view.  The need to extract a wide, sometimes slightly out-of-focus 
spectrum, coupled with the limited amount of sky covered by each 
slitlet, prevented us from using any standard reduction method.

To reduce our spectra, we began by debiasing the raw CCD frames, and 
extracting the data region of each slitlet.  We then created a normalized
flatfield frame for each slitlet using observations of both the interior dome
screen and the twilight sky.  To do this, we first removed the effects of
vignetting from our dome flats by fitting a high-order spline to the data,
and then dividing the flatfield by this function.  We applied this
normalized flatfield to spectra of the twilight sky, and examined the 
resulting image for residual structure in the spatial direction.  If any 
structure was apparent, we fit it with a low-order polynomial, 
and re-normalized the dome flat to remove the effect.  Once the final,
normalized dome flat was calculated, we used it to flatten all spectra
taken through the slitlet.

After flatfielding the data, we carefully examined the sky spectrum
surrounding each PN, paying particular attention to the nebular emission
lines near H$\alpha$.  As detailed by Ciardullo \etal (1988) and Jacoby,
Ford, \& Ciardullo (1985), the bulge of M31 contains a series of irregular
emission features, whose [N~II], [S~II], and H$\alpha$\ flux can be as
bright, or brighter than, the target planetary.  This greatly  confuses
the issue of sky-subtraction, for a single knot of emission within a
slitlet can cause the background at these wavelengths to be severely
overestimated.  To reduce this problem, we inspected the data of each
slitlet, and identified any anomalously bright emission regions along
our slit.  If an emission knot was detected, its presence was noted and
that region of the slitlet was excluded from the analysis.

Once the inspection was complete, we proceeded to extract the object
spectrum from each slitlet.  To do this, we used our standard star
observations to trace both the shape and the varying width of the
spectra produced in each slitlet.  We then applied these parameters to
our program objects, using the bright [O~III] $\lambda 5007$ line as a
reference to define the exact position of the PN within the slitlet.
Our extraction program defined three regions: an object region, which
changed in size along the spectrum and included $\sim90$\% of the
object counts, and two sky regions, one on either side of the object
(and separated from the target spectrum by $\sim 3$ pixels).  Using the
sky regions, we estimated the sky value underlying each pixel in the
object region, either by adopting a median sky value, or, for those 
PNe close to the nucleus, by fitting a low-order function to the data in 
the spatial direction.  The sky was then subtracted from the object pixels, 
and the object counts summed to produce a one-dimensional, sky-subtracted
spectrum.

Following the sky-subtraction, the data reduction continued in a more
standard manner.  By dividing our exposure into $\sim 12$ separate 30-min
frames, we were able to identify cosmic rays via a sigma-clipping algorithm:
any point that was more than $\sim 10 \, \sigma$ from the median was deleted
from the analysis. A series of He-Ne-Ar comparison arcs taken throughout
the night served as the wavelength calibration.  Each individual spectrum
was linearized, corrected for atmospheric extinction, and flux-calibrated
using the comparison arcs and the observations of the spectrophotometric
standard stars.  Finally, the $\sim 12$ spectra of each PN were combined,
using the number of counts recorded in the [O~III] $\lambda 5007$ line
to weight the averaging.  The final reduced spectra are shown in
Figure~1 (bulge PNe) and Figure~2 (disk PNe).

\subsection{Reduction to dereddened line ratios}

The integrated flux in a line was derived using the Gaussian line
fitting and deblending tools in the SPLOT task of IRAF{}. Line strengths
in each spectrum were measured at least twice, to assess the subjective
uncertainty in placing the continuum, which is one of the dominant sources
of error
for the faint lines. Typically, we found that the measurement uncertainty
is roughly 2\% of the strength of H$\beta$, except for the region near 
$\lambda 3727$, where the system sensitivity is dropping quickly.  In
this region, the errors depend on the background signal relative to the
PN signal, and are typically magnified by factors of 3 to 10.

The spectrum of one PN, CJFN 9, was so weak that no usable information
could be extracted. We dropped that PN from our sample. The remaining
15 PNe have line strengths after dereddening (Cardelli, Clayton, \&
Mathis 1989)  as given in Table~2.  The logarithmic extinction at H$\beta$,
$c$, derived from the Balmer decrement, $I({\rm H}\alpha)/I({\rm H}\beta)$, is
also listed in the table.  The uncertainty in this ratio is typically
about 5\%, such that $\delta c \sim 0.06$.

Table~2 also lists the absolute extinction-corrected flux, $L_{5007}$,
of each planetary.  These numbers were derived by taking the magnitudes
tabulated by CJFN, and scaling them to an M31 Cepheid distance of 770~kpc
(Freedman \& Madore 1990).  As discussed in the next section, the nebula 
luminosities provide an important constraint on the models, and help
define the central star luminosity and mass.

\section{Analysis Using Photoionization Models}

As can be seen from Figures~1 and 2, the quality of some of the PN spectra
are excellent: lines as faint as 10\% of H$\beta$ can be measured accurately. 
On the other hand, many of the spectra are missing lines such as 
[S~II] $\lambda\lambda 6716,6731$ and [O~III] $\lambda 4363$, which are
necessary to derive electron densities ($N_e$) and temperatures ($T_e$).
In order to proceed with an abundance analysis in the latter
cases, one can adopt reasonable values for these parameters (e.g., $T_e =
10,000$K; $N_e = 4,000$). Or, since the more important parameter is $T_e$,
one can estimate an upper limit for $I(\lambda 4363)$, infer an upper limit 
for $T_e$, and consequently, derive a lower limit for the abundances of the 
collisionally excited elements. This latter approach was adopted by SRM in 
their analysis of extragalactic PNe.

Rather than use this approach, we have chosen to use the CLOUDY 90.04
photoionization code (Ferland \etal 1998) to indirectly deduce the viable
range of elemental abundances using whatever emission-lines strengths
(or upper limits) that are available.   Though the technique is time
consuming, and not necessarily suitable for large volumes of data (such as
that of the SRM study), the method does have the advantage of allowing
us to investigate the properties of both the nebula and the central
exciting star. An intangible asset of using photoionization modeling
is that the process focuses one's thinking on the physical properties
(radius, luminosity, dust content, central star temperature) of each PN
as a real entity rather than simply a column of line strengths.

\subsection{Philosophy of the modeling process}

Generally, for our sample of PNe, there were too few lines to fully
constrain all aspects of a photoionization model. Consequently, we modeled
each object with the simplest set of parameters possible.  Complexities
were added only if a viable model could not be built, where we defined
a viable model as one in which a) the predicted emission-line ratios
matched the observed line ratios to within 2\% of $I({\rm H}\beta$), and 
b) the total
[O~III] $\lambda 5007$ flux of the model matched the observed, dereddened
flux.  As a result, our basic model consisted of a hot
black-body central star centered in a spherical nebula having uniform
density and containing dust grains typical of most PNe.  In two cases
the uniform density assumption failed, and a multizone model was
needed, in six cases dust was omitted from the nebula, and, in one
difficult case (CJFN 478) the dust fraction was enhanced well above
that seen in normal nebulae.  It was never necessary to relax either the
black-body or the spherical nebula assumptions. This condition, though,
should not be confused with any demonstration that the central stars
are black-bodies, or that the nebulae are spherical.

The procedure used to iterate toward a solution went roughly as follows.
The nebula was first initialized to a set of abundances typical of
a Galactic PNe, and to a radius large enough (usually about 0.3 pc)
so that the nebula was thick to all ionizing radiation.  (To compute
this latter condition, we used the [S~II] $\lambda\lambda 6717,6731$
density measurement when it was available, otherwise we assumed $N_e =
4,000$~cm$^{-3}$.)  The central star temperature was then varied until
the helium line ratios were approximately correct.  For the three PNe
where the helium lines were not reliably detected (CJFN 75, 455, and
470), the central star temperatures were constrained by the strength of
$\lambda$5007 and the upper limits to the helium lines.  The central
star luminosity was adjusted to yield an approximately correct value
for the absolute $\lambda 5007$ flux.  If necessary, the nebula radius
was then reduced to lower the line fluxes of the low ionization species
(e.g., [O~II], [N~II], [S~II]), and a first adjustment was made to the
abundances in order to improve the predicted line ratios.

At this point, secondary adjustments were made to the central
star temperature and luminosity, the nebula radius and density, and to the
abundances.  In some cases, the amount of grains in the nebula was 
modified as well, in order to achieve a better match with the helium 
and oxygen lines.  Finally, for a few objects, it became necessary to 
alter the abundances of elements for which no lines were available.  The most
important of these elements is carbon, whose presence can have a 
significant effect on the overall cooling efficiency of the nebula.  By
adjusting the amount of nebular carbon (while keeping it reasonably
consistent with the metallicity defined by the other elements), 
we were able to fine-tune the nebula's electron temperature and improve
the corresponding line ratios.  

Generally, our nebular modeling approached convergence after 10-20
iterations, although, for a few of the best observed objects, another
20 iterations were needed to achieve a good match to all the line
ratios\footnote{When a doublet is present, such as the $\lambda\lambda
4959,5007$ lines of [O~III], the model is forced to match the brighter
line of the pair.}.  Table~3 summarizes the parameters describing the
15 nebulae.  Included in the table are the derived values for the nebula's
electron density, temperature, and radius, the central star's effective
temperature and luminosity, and abundance values for helium, oxygen,
nitrogen, neon, sulfur, and argon.  Because of its importance
to the nebular modeling, the assumed carbon abundance is also reported,
although no lines of carbon were directly observed.  
The electron temperatures and densities are volume-averaged quantities.

The central star mass also is included in Table 3.  These masses
have been computed by interpolating the central star's location in
the HR diagram onto the grid of hydrogen-burning evolutionary tracks of
Bl\"ocker (1995) and Sch\"onberner (1983).  Thus the mass estimates
carry the additional uncertainty associated with the applicability of
these stellar evolutionary models.  In particular, if any of the PNe
considered in this survey are helium burners, then their derived mass
estimates may be significantly in error.

\subsection{Accuracy of the models}

In nearly every case, our models were able to reproduce all the observed
line ratios to within $\sim$0.02 of the measured value (see Table
2). This is well within the estimated measuring errors of $\sim 5\%$
(systematic) plus $>3\%$ (random). For weak lines ($<20\%$ of I(H$\beta$))
in well-observed objects, the random errors are typically $<20\%$. Obviously,
for fainter objects, the errors are worse.

In the two cases where a multizone density model was required, single
zone models yielded abundance values that differed by $<0.3$ dex (usually
$<0.1$ dex). Abundance uncertainties of this magnitude, though, are large.
Generally, when the data quality is good to excellent, our abundance
measurements are constrained to $< 0.1$ dex. This is especially true for
oxygen, neon, and helium where, typically, the models are constrained
to $\sim0.04$ dex ($\sim 1 \sigma$). 
Nitrogen and sulfur abundances are more uncertain
since these elements have only one visible ion, whose contribution is
dependent on the geometry of the nebula.  (Although the [O~II] 
$\lambda 3727$ to [O~III] $\lambda 5007$ ratio does provide an independent
measure of this geometry, this constraint is only applicable under the
assumption of sphericity.)  We estimate that our N/H and S/H are
accurate to $\sim 0.3$ dex.  Argon is modeled with good stability, but
depends strongly on an accurate measurement of the [Ar~III] $\lambda$7135 line.
Because this feature is generally not well observed, our Ar/H values also
are uncertain at $\sim0.3$ dex.

With respect to the central star properties, the stellar temperatures are
generally constrained to within $\sim 5,000$~K when the He~I and He~II
lines are measured with good accuracy (6 PNe), to $\sim 10,000$~K when
only He~I is measured well (4 PNe), and to $\sim 20,000$~K otherwise.
Similarly, the stellar luminosities are constrained by the [O~III]
$\lambda 5007$ fluxes, which are well-determined from photometric
observations.  Second order factors include the correction for extinction,
the adopted fraction of dust in the nebula, and the derived optical
thickness to ionizing photons (which comes from the ratio of [O~II]
$\lambda 3727$ to [O~III] $\lambda 5007$).  Based on our experiments
with near-acceptable models, these effects generally introduce an
uncertainty of $\sim 0.1$~dex on the central star luminosity, although
the luminosities for PNe with low quality spectra may be in error by up
to 0.3 dex.

Since nearly all of the bright PNe considered in this paper are on the 
horizontal portion of their evolutionary tracks, the central star masses 
are not significantly affected by errors in temperature. Luminosity errors,
however, do propagate into the mass determinations. For typical
central stars with $M \sim 0.6$ M$_\odot$, the error in mass is always
$< 0.07$ M$_\odot$, and usually less than $\sim 0.02$ M$_\odot$.

In the next section, we discuss another aspect of the accuracy of
the models.

\subsection{Uniqueness of the models}

Several authors (Jacoby \etal 1997; van Hoof \& Van de Steene 1997)
have described attempts to probe the robustness and uniqueness of
PN photoionization models.  The results are encouraging, but unless
there are many emission-lines from many different ions, some level of 
uncertainty always remains.

For two nebulae having high central star masses (CJFN 27 and CJFN 1),
we attempted to derive models with core masses more in line with
the other nebulae.  These attempts were not successful; in fact,
the masses could not be changed by more than $\sim0.06$ M$_{\odot}$ from
the initial converged value.  Our models for these objects may therefore
be unique.  

For the two nebulae requiring multizone density models, we pursued
single zone models until it became apparent that such a model would never
succeed.  Hence, for these objects, single zone models were ruled out.
Unfortunately, with the added flexibility provided by a less constrained
density distribution, the results for these models may be less secure
than those of the remaining nebulae.

Observations of the nebula CJFN 27 were matched with a very high density
nebula ($N_e = 89,000$ cm$^{-3}$).  To check this, we attempted to 
model the nebula with a lower, more typical density ($N_e = 9500$ cm$^{-3}$). 
This model succeeded in meeting the observational constraints, but in
order to maintain a high electron temperature, the carbon abundance had
to be reduced by a factor of four, and the dust component had to be increased
to 2.4 $\times$ the typical PN fraction. This alternative model also
demonstrated marginal stability in the sense that a small change in {\it any\/}
parameter rapidly drove the solution away from many of the observational
requirements.  More importantly, while the elemental abundances in the
alternate model did change, the worst case changes were only +0.26 dex 
(for nitrogen) and $-0.21$ dex (for oxygen).

In addition to these experiments, we pursued one additional test of
uniqueness.  We attempted to build a model of the reasonably well-observed
PN FJCHP 51 with an oxygen abundance 0.3 dex lower than the nominal value,
but with no constraints on the strength of the nebula's [O~III] $\lambda
4363$ line.  In other words, we simulated the case where this weak, but
important, line is lost in the noise.  The test failed in the sense that
we were not able to create a successful model in which O/H was more than
0.16 dex below normal.  Moreover, in order to achieve even this modest
level of reduction, the nebula's carbon abundance had to be significantly
(and perhaps unrealistically) lowered by 0.9 dex 
to force a high electron temperature; under these conditions, the [O~III]
$\lambda 4363$ line would be $2 \sigma$ stronger than that observed.
From this, we conclude that the maximum deviation in derived oxygen
abundance is within the range of the observational errors.

A full investigation into the uniqueness of photoionization modeling is
beyond the scope of this paper.  Our experience, though, indicates that
it is difficult enough to get {\it any\/} model to converge to the
appropriate line ratios, let alone to devise successful models with vastly
different properties.

\subsection{The Balmer decrement}

During the extinction and reddening correction phase of the reductions,
we adopted a standard value of 2.85 for the Balmer decrement ratio of
$I({\rm H}\alpha)/I({\rm H}\beta)$.   This, however, is a simplification:
the true Balmer decrement can range between 2.7 and 3.1, depending
on plasma conditions (cf. Ferguson \& Ferland 1997; ~Osterbrock 1989;
Ferland \etal 1998).  Thus, to be fully self-consistent, we should have
revised our extinction estimates based on the derived values for the
electron temperatures and densities and then recomputed our nebular models.
This iteration was not performed.  For 11 of the 15 PN considered here,
the derived ratio for the Balmer decrement was within 1\% of the adopted
value, and thus the additional iteration was not needed.  For the
remaining PNe, $I({\rm H}\alpha)/I({\rm H}\beta)$ ranged from
2.76 to 3.06, or $\pm 4$\%.  Since the systematic errors in the line
measurements (and uncertainties in the physics of the hydrogen atom)
are comparable to, or worse than, this extreme range, we decided that the
small error introduced by this effect was acceptable.

\subsection{Comments on the nebulae}

{\it CJFN 1:} [$37''$ from the nucleus] Although this PN is close
to the nucleus, it is very luminous: only 0.5 mag down the PNLF{}.
Consequently, the spectrum has a reasonably high signal-to-noise.  Our
models demonstrate that this nebula is almost completely optically thick,
with a radius that is 97\% of the optically thick value.  A very high
central star temperature is required to meet the strong He~II line strength.

We note that RSM also observed this object, though with lower 
signal-to-noise.  They report similar line ratios, at least for those
lines that they were able to measure.  Whatever differences exist can be
attributed to the somewhat lower quality of their spectrum. 

{\it CJFN 9:} [$8''$ from the nucleus] This object is very close to
the nucleus of M31 and located on top of significant diffuse emission. The
combination of a bright stellar background and bright emission from the
local interstellar medium
rendered this spectrum unusable (\ie the PN Balmer lines could
not be measured). No results are presented for this object.

{\it CJFN 23:} [$68''$ from the nucleus] There are no low excitation
lines visible in this PN, but this may be due to the poor 
subtraction of M31's diffuse emission.  Without the low ionization species,
there is no way to constrain the ionization structure of the nebula with 
confidence.  The red region of the spectrum suggests that the sky
subtraction went well and that the object is a high excitation nebula with 
very weak lines of [N~II] and [S~II].  The blue [O~II] lines, however, are 
over-subtracted, but the implications of this are difficult to evaluate 
because of the instrumental defocussing.  In order to model the upper limits 
on the [N~II] and [S~II] lines ($\sim 0.2$ of $I({\rm H}\beta)$), a nebula 
radius of 0.061 pc is required, which is 90\% of the optically thick value.  

Alternatively, N/H and S/H could be reduced by making the nebula more 
optically thick and enlarging its radius. If the nebula is made less
optically thick, N/H and S/H can be increased substantially.  However,
in this case, O/H also would have to be increased to meet the nebular flux 
constraint.

This PN also was observed by RSM; their line ratios agree approximately
with ours, but neither data set has a high signal-to-noise ratio.

{\it CJFN 27:} [$152''$ from the nucleus] This PN spectrum is excellent,
as might be expected from a very luminous PN (0.6 mag down the PNLF).
Yet, the low ionization lines are extremely faint, with the [N~II] lines
being the only ones measured with confidence. [S~II] is marginally
present, but the ratio of $I(\lambda 6717)/I( \lambda 6731)$ is at
the low density limit, which is physically unrealistic for a bright
PN{}. Presumably, if the [S~II] intensities are correct, they derive
from a very tenuous outer region beyond the main body of the nebula.
This object also exhibits very strong [O~III] $\lambda 4363$ emission;
this implies a substantial amount of dust, since this helps the nebula
attain a high electron temperature.  The large number of lines and
ionization states observed, coupled with the excellent diagnostic provided by 
$\lambda 4363$, make this one of our best constrained models.  Nevertheless, 
the very high density deduced for CJFN 27 suggests that an alternative model 
might be more likely.  Unfortunately, our attempts to devise a low density 
model for this object were only marginally successful (see \S 3.3). The 
nebula is optically thick.

{\it CJFN 29:} [$132''$ from the nucleus]  Because the [S~II] line ratio
is at the low density limit, this planetary must contain a significant
amount of material at low density.  However, the high [O~III] flux
of the object requires a much higher density; otherwise, in order to
produce this flux, the nebula would have to be extremely large and massive.
This discrepancy drove us to consider a two-zone density structure for this
object.

During the modeling process, we noted that the observed [S~II] ratio
$I(\lambda 6717)/I(\lambda 6731)=1.76$, which suggests a density close to
zero.  Since this is unphysical, it implies that there is an error of
at least $\sim 25$\% in our measurement of these weak lines; presumably,
this is due to the poor subtraction of the diffuse ISM present in M31's bulge.
To deal with this problem, the density of the model's outer zone (which
starts at a radius of 0.054~pc and terminates at 0.462~pc, or 81\% of the 
optically thick radius) was arbitrarily set to a density of 280 cm$^{-3}$.  
This forced the sense of the [S~II] lines to agree with the observations 
(i.e., to guarantee that $I(\lambda 6717) > I(\lambda 6731)$, as seen even 
in the raw data).  The models were then required to match only the brighter of
the [S~II] lines, while the weaker line floated to a value indicative of
the adopted density in the outer zone.  Meanwhile, the density of the
inner zone was arbitrarily set to 7000 cm$^{-3}$.

We note that the high value for [O~III] $\lambda 4363$ presented a minor 
problem for the models.  To offset the very high abundances derived for the
object, dust was needed at 90\% of the nominal value for PNe.  Another
oddity is that due to its high abundance ([S/H] = +0.5), cooling via
sulfur appears to be very important for this object.  On the other hand,
this result may be an artifact of the weak sulfur lines and difficult
background subtraction.

A single-zone model was also attempted with limited success.   For this
model, we left the [S~II] line ratio unconstrained but matched 
the total [S~II] flux of the object.  The result was that none of the 
abundances changed by more than $\sim 15$\% (0.06 dex) except for nitrogen, 
which decreased by 0.2 dex.  The central star luminosity, however,
dropped by 35\% since the single zone nebula was more optically thick than
the two-zone model.

This nebula is unusual in yet one more way. It is the best observed
object in common with RSM, but a major discrepancy exists for the relative 
strength of [O~III] $\lambda 5007$. Our value of $14.8 \times I({\rm H}\beta)$
is significantly different than the value that they report ($22.2 \times
I({\rm H}\beta)$).  We and RSM have attempted to resolve this disagreement
but have not identified any problems in either of our data sets.
We expect our observations to be accurate to $\sim 3\%$.  The discrepancy
is more mysterious when we consider that the line ratios for two fainter
PNe, CJFN~1 and CJFN~23, are in reasonable agreement.  One possibility is that
CJFN 29 is variable on a short timescale, though we note that the RSM data
were taken only 13 months before our observations. In support of this 
speculation, we do see very good agreement in the derived elemental
abundances ($\Delta$ [O/H] = 0.04).

{\it CJFN 33:} [$219''$ from the nucleus] The spectrum for this PN is
of moderate quality, but no low ionization lines are present. The radius
was set to the maximum that did not produce [O~II] $\lambda 3727$
above the observed limit of $\sim 0.25$ of $I({\rm H}\beta)$.  Beyond the UV,
most lines as faint as 10\% of $I({\rm H}\beta)$ are well measured; the lone
exception is He~I $\lambda 5876$, since the velocity of CJFN~33 places the
line in the Na~I sky feature.  Perhaps this is why the derived helium
abundance is abnormally low ($N({\rm He})/N({\rm H}) = 0.05$).  
With $I(\lambda 5876) = 
0.06 \, I({\rm H}\beta)$, this low abundance is inevitable and is independent 
of any modeling ambiguities. The model is optically thin, with the outer
radius at 83\% of the optically thick value.

{\it CJFN 75:} [$78''$ from the nucleus] Only the He~I line ($\lambda
5876$) presented a problem in developing a model for this nebula. The
measured value of 0.76 of $I({\rm H}\beta)$, though, is very likely to be
an artifact of poor background subtraction. The helium abundance was
arbitrarily chosen to be 0.148 which is a high value, but not unusually so. The
strength of the [O~II] lines sets the outer radius to be at 97\% of the
optically thick value. However, other than the usual requirement for absolute
nebular flux, the model is not tightly constrained.  Still, we believe the
O/H and Ne/H abundances are valid.

{\it CJFN 125:} [$250''$ from the nucleus]  Although this bulge PN is 
relatively far from M31's nucleus, it is intrinsically faint (1.6 mags down 
the PNLF) and thus, its spectrum is quite noisy.  The lines of [O~III] 
$\lambda 4363$ and [S~II] $\lambda\lambda 6717,6731$ were not detected, 
although lines as faint as 10\% of $I({\rm H}\beta)$ are measurable.  Given 
the limited number of lines, a viable model with no dust was found easily. 
The model is optically thin, with the outer radius at 83\% of the optically 
thick value.

{\it CJFN 177:} [$206''$ from the nucleus]  Despite being 1.8 mags
down the PNLF, the spectrum of this nebula is very good, and it displays a
relatively rich variety of ionization states.  Thus, the object can be modeled
with enough constraints to delimit abundances fairly well. Lines
as faint as 5\% of $I({\rm H}\beta)$ are measurable; [O~III] $\lambda 4363$
may have been detected at 2\% of $I({\rm H}\beta)$.  The model fits all lines
(including the limit on $\lambda 4363$) with a nearly optically thick
nebula: the derived outer radius is 98\% of the fully optically thick value.

{\it CJFN 179:} [$234''$ from the nucleus] This PN is intrinsically
faint, 2.2 mag down the PNLF{}. Furthermore, it falls on a region
of diffuse emission, leading to complications in background subtraction. Lines
as faint as 12\% of $I({\rm H}\beta)$ are measurable, but the important
diagnostic lines are not seen. Given the limited number of lines, a
viable model with no dust was found easily. The model is optically thin,
with the outer radius at 81\% of the optically thick value.

{\it CJFN 455:} [$137''$ from the nucleus] This PN is very faint and
close to the nucleus. It suffers more than any other from background
subtraction errors, as evidenced by the negative (i.e., over-subtracted)
lines of [O~II], [N~II], and [S~II]. Since the ionization structure could
not be derived from the very limited number of line ratios, our model was
defined to have a radius at 90\% of the optically thick radius, a number
typical of many of the M31 PNe. With only lines of [O~III] and [Ne~III]
measurable, this nebula model is very poorly constrained. However, the
[O~III] luminosity serves to set the luminosity of the central star which,
in combination with any likely temperature, allows an estimation of
the central star mass.

{\it CJFN 470:}  [$101''$ from the nucleus] This PN has the faintest
apparent $m_{5007}$ of the sample, 3.0 mag down the PNLF, and it has a
relatively low excitation.  Due to its faintness, few lines are
measured well, and the background subtraction and contamination errors
from the diffuse ISM of M31's bulge are significant. Abundances and
stellar parameters are indicative, but not accurate. Since the error in
$I({\rm H}\beta)$ alone is $\sim 20$\%, no line ratios are more accurate that
this; nevertheless, a model with the standard 2\% error was devised. The
model required a central low density (140 cm$^{-3}$) region extending out
to 0.50 pc; beyond that and to a radius of 0.62 pc, a density enhancement
N$_e$ $\sim$ 250 cm$^{-3}$ is required. Although there are very few
constraints to the model, two could not be matched.  First, the He~I
line at $\lambda 5876$ is so strong that it must be an artifact; the
spectrum is very noisy in this region and the apparent line is blended
with sky emission. Second, the weaker [S~II] line ($\lambda 6731$) could
not be matched within the usual tolerance (0.02 of $I({\rm H}\beta)$), but was
reproduced at a strength of 1.66 $I({\rm H}\beta)$, 13\% lower than observed.
Given the large diameter of this nebula (1.24 pc, or $0\parcsec 33$), the 
object should be easily resolvable with the {\sl Hubble Space Telescope\/}
if the model is correct.

{\it CJFN 478:} [$137''$ from the nucleus] This nebula presented a number
of problems in the modeling.  First, the [O~III] $\lambda 4363$ line is
measured to be moderately strong at 0.042 $\times$ 
$I(\lambda 5007)$: this indicates a high electron temperature, 
$T_e = 22,000$~K{}.  However, to attain this temperature, the carbon
abundance (and its cooling contribution) must be decreased significantly,
and the amount of dust grains must be increased to five times the normal
PN dust-to-gas ratio.  The high dust factor serves two roles: it reduces
the nebular cooling by depleting the metals onto grains,
and it provides an additional heating source via photoelectric emission
(Borkowski \& Harrington 1991).  Unfortunately, the addition of so much
dust also reduces the emitted luminosity of the nebula, forcing the central
star to be intrinsically more luminous and massive ($M > 0.9$ M$_{\odot}$).
Given the halo-like abundances of this nebula, this seems unlikely.
The alternative is to suppose that the flux from [O~III] $\lambda 4363$
has been overestimated; since the signal-to-noise of the line is not high
(the line is only measurable to $\sim 30\%$ accuracy), this possibility is
quite viable.  On the other hand, we also have a marginal detection of the
[N~II] $\lambda 5755$ line, which provides additional evidence for a high 
electron temperature.  

Because of this confusion, the results presented in Table~3 represent those
of a compromise model that trades off between a high dust content and
a high central star luminosity.  In the model, the dust is enhanced by
a factor of 3, the central star mass is $\sim 0.88 M_{\odot}$, and 
the strength of [O~III] $\lambda 4363$ is assumed to be 30\% smaller
than actually observed. This model has a radius of 0.36~pc, only 5\%
short of the optically thick value.

A similar uncertainty arises when we consider the strength of [O~II]
$\lambda 3727$.  The high ratio of [O~II] $\lambda 3727$ to H$\beta$\ 
(1.05) is  easily reproduced with a two-zone density model.  In fact,
the compromise model yields $I(\lambda 3727) = 1.12$; in the model,
the inner nebula (radius $<$ 0.018 pc) has a moderately high density
(11,000 cm$^{-3}$), while the outer region (extending to a radius of 0.36~pc)
has $N_e \sim 460$ cm$^{-3}$. Again, however, the [O~II] line suffers from
an added uncertainty associated with the contribution of irregularly
distributed diffuse emission in M31's bulge.  Moreover, in this part
of the spectrum, instrumental factors (low sensitivity, defocus of the
spectrum in the camera) combine to lower the signal-to-noise of the
measurement.  Consequently, a single zone model does as well as the 
two zone model if the strength of the [O~II] line is allowed to drop
by almost a factor of two, to 0.65 $I({\rm H}\beta)$. This degree of error
is unlikely, so we adopt the compromise dual zone model.  The differences
in the abundance determinations between the single and dual zone
models are always $<0.15$ dex.

{\it FJCHP 51:} This PN and the other two outer PNe were far easier to
observe than any of the bulge objects because of the much lower background
levels.  Consequently, the models for these three objects have much better
constraints than those for their bulge counterparts. For example, both
the [O~III] $\lambda 4363$ and [N~II] $\lambda 5755$ temperature sensitive
lines, as well as the density sensitive lines of [S~II], were measured in
this nebula. An excellent match to the observed line ratios was obtained
with a radius at 93\% of the optically thick value.

{\it FJCHP 57:} This intrinsically luminous PN exhibits relatively low
abundances. The spectrum is very good and has the unique property that
features from the central star are visible. The Wolf-Rayet signatures at
$\lambda 4650$ from (N~III, C~III, and C~IV) and $\lambda 5806$ (from
C~IV) are seen easily. The relative strengths of these lines, plus the
absence of other Wolf-Rayet indicators, suggest that this central star
is approximately a WC~6 class object (van der Hucht \etal 1981). The
temperature calibration of Tylenda, Acker, \& Stenholm (1993) then gives
$T_{eff} \sim 50,000$~K, in excellent agreement with the temperature
of 58,000~K needed by the photoionization model to match the observed
strength of the helium line at $\lambda 5876$. The nebula is nearly
optically thick, with a radius at 98\% of the optically thick value.

{\it FJCHP 58:} This is a high excitation PN with no evidence for lines
of [O~II], [N~II], or [S~II].  This lack of emission is significant, since
the spectrum has a fairly high signal-to-noise, and non-UV lines as weak as
4\% of H$\beta$\  should be seen.  (For [O~II], the limit for detection is
closer to 20\% of H$\beta$.) The geometry was defined to meet these limits,
leading to a radius that is 86\% of the optically thick value. The relatively 
low luminosity of the nebula is responsible for the low central star 
luminosity and mass.

\section{Results}

\subsection{Abundance comparison with ICF method}

In the 8 cases where plasma diagnostics allow a direct comparison between
nebular models and measurements based on ionization correction factors
or ``ICFs'' (e.g., Alexander \& Balick 1997), we generally find good
agreement in the abundances.  The summary of the differences between the
CLOUDY and the ICF abundances (as formulated by Alexander \& Balick 1997)
is given in Table~3.  On average, the photoionization method yields
slightly higher abundances, typically 20\% (0.08 dex).  Moreover, for
the worst case nebula, CJFN~27, if the alternative low density model is
adopted, the agreement with the photoionization models is improved, but
CLOUDY still yields higher abundances by 15\% (0.06 dex).  Since nebula
CJFN~27 is responsible for the most extreme difference between the two
abundance techniques, the latter estimate is more typical of the sample
as a whole.

The abundance differences between the CLOUDY photoionization results
and the values derived from the ICF method arise from two primary
parameters: $T_e$ and $N_e$.  First, the model's estimate of $T_e$
assumes a limited uncertainty in $\lambda 4363$; in contrast, the ICF
method accepts the observed strength of $\lambda 4363$ as exact. Thus,
within the observational uncertainties, the two methods will yield
the same abundances if $T_e$ is the dominant source of difference.
Second, the density suggested by the [S~II] lines is very uncertain, both
because these lines are usually weak in high-excitation planetaries, and
because the low-excitation diffuse emission in M31's bulge complicates
the background subtraction.  (Indeed, as described above, the implied
[S~II] densities can occasionally be unphysical.)  Because the
ICF method assumes that the [S~II] ratios are exact, errors in their
measurements will affect ICF abundances more than model derived values.

Based on the above arguments, we conclude that even when line ratios
are available for ICF measurements, there still is a small advantage
in accuracy obtained by deriving abundances of extragalactic PNe via
photoionization models.  Alexander \& Balick (1997) also concluded that
the ICF methods generally are adequate, but could suffer additional
inaccuracies under certain conditions.

\subsection{Abundance comparison with SRM}

We observed 3 PNe in common with the sample of SRM (CJFN 1, 23, and
29). Given that the observations, reductions, and analysis are completely
independent and utilize very different approaches, these objects provide a
valuable external test of the errors in the abundances.   Immediately
obvious is the fact that the SRM abundances for helium in all three
objects are unrealististically low, typically He/H $\sim 0.015$.  Our
values are closer to those observed in Galactic and Magellanic Cloud PNe.  
For oxygen, however, the agreement is excellent: the logarithmic difference 
in the two measurements is $\Delta {\rm O/H} = -0.013 \pm 0.05$ (ours minus 
SRM).  The results for neon are mixed, with $\Delta {\rm Ne/H} = 0.00 \pm 
0.18$.  Although there is no systematic offset between the abundance 
determinations, the dispersion between the two neon measurements is large.

The fact that the neon measurements exhibit a large amount of scatter should
not be too surprising.   The abundance of neon is based on a single,
relatively weak line of [Ne~III] at $\lambda 3869$.  Thus, with only one
ionization state available, abundance measurements carry a larger than
normal uncertainty, especially for ICF-based values.  In addition, 
the [Ne~III] line lies well into the UV, where system sensitivities are poor
and where atmospheric dispersion effects in uncompensated observations have
the greatest impact.  (Our observations were made using the atmospheric
dispersion corrector built into the 4-m telescope.)  Nevertheless, it is
comforting to see that there appears to be no serious zero-point errors between
the measurements.

\subsection{Abundance correlations}

Further confidence in the derived abundances of O and Ne is provided by
their very tight correlation (first noted by Henry 1989). 
This is displayed in Figure~3.  It is clear from
the figure that our PNe sample exhibits a wide range of oxygen abundances 
from 0.05 solar to marginally super-solar with a clump of objects near 
O/H $\sim$ 8.35.

Also plotted in Figure~3 are the abundances measured by SRM{}. Their
PNe follow exactly the same locus as the 15 objects presented here, but
exhibit higher median abundances.  (This is due to a selection effect in 
luminosity; see \S 4.4.1.) There are, however, 2 PNe in our sample that lie off 
the main locus of points -- FJCHP 57 and CJFN 455.  FJCHP 57 is a high velocity
planetary that has generally low abundances reminiscent of 
halo PNe in the Milky Way. Henry (1989) has found that Galactic halo
planetaries can deviate from the main locus in the O--Ne correlation. If 
FJCHP 57 is really a halo PN in M31, it is projected on a disk field and is
therefore a fortunate find.  CJFN 455, on the other hand, is a very faint 
object with a marginal measurement for neon.  Thus, it is not significant that 
it falls off the O--Ne relation.

In Figure~4, we show the correlation between N/O and N/H{}. Even with the
limited data, it is clear that the trend seen by Henry (1990), and shown
by the dashed line, is followed by the M31 PNe. The slope of the trend
is 0.96, which is close enough to unity to suggest that variations in
N/O are due almost entirely to variations in N, independent of O{}. That
is, nitrogen is not being manufactured at the expense of oxygen.  This
conclusion is supported by the absent, or possibly weak, anti-correlation
between N/O and O/H (Figure~5). A similar weak anti-correlation has been
reported by others for Type~I PNe (see Henry 1990 and references therein),
but this was not seen in the large sample of Kingsburgh \& Barlow (1994).

The correlation seen in Figure~4, though, could be
due entirely to large observational errors in N/H, since those would
affect N/O similarly. This explanation, however, would also have to apply to
Henry's sample unless one accepts a coincidence of nature in which our
(accidental) correlation matches Henry's real one.

Figure~5 presents a variety of relationships (or lack thereof) between N/O
and He/H, O/H, extinction, $m_{5007}$, and core mass.   Four of the 
objects have $\log$ (N/O) $> -0.3$ and He/H $> 0.125$; according to the
criteria of Peimbert \& Torres-Peimbert (1983), these are Type~I
planetaries.  This is somewhat surprising: in the Galaxy, Type~I PNe are
presumed to derive from younger, more massive progenitors, yet three of
these putative Type~I PNe are located in M31's bulge.  Moreover,
if, in fact, our Type~I PNe do arise from massive progenitors,
we might expect the high N/O objects to be brighter than their low
N/O counterparts, since their central stars would be more luminous and
more massive (cf.~M\'endez \etal 1993).  The anti-correlation between N/O and 
$m_{5007}$ (Figure~5), a consequence of added nitrogen competing with
oxygen to cool the nebula (Jacoby 1996), shows that the opposite is true.
In fact, the lack of correlation between N/O and core mass argues
that although our PNe with high N/O may have abundances typical of 
Type~I PNe, their
central stars come from an older, less massive population.  Probably
some, or all, of these objects are not true Type~I PNe, but exhibit
high N/O and He/H because their progenitors were enriched (see also SRM).
Curiously, our 
Type~I candidates tend to be among the least dusty PNe in our sample.  But
this could be a selection effect: a highly extincted object would
not have been detected in the CJFN survey.

No other clear trends between abundances are seen. There may be a weak
positive correlation between Ar/H and O/H, but the errors in the argon
line ratios are large enough to render any such trend suspect.

\subsection{Other correlations}

\subsubsection{abundance and luminosity}

Given that the strength of the [O~III] lines is proportional to the number
of oxygen atoms, one might expect that, all things being equal, an
oxygen-rich PN would be more luminous in $\lambda 5007$ than an oxygen-poor 
object.  However, all things are not equal.

With more oxygen (and presumably more of other elements as well), nebular
cooling is more efficient.  This lowers the nebular electron temperature
and decreases the collisional excitation that drives the [O~III] lines.
As a result, the nebular luminosity in $\lambda 5007$ is expected to rise 
roughly as the square root of O/H rather than O/H directly (Jacoby 1989; 
Dopita \etal 1992).  The effects of metallicity are then 
complicated further by the PN progenitor's evolution prior to its departure 
from the AGB{}. High metallicity stars are expected to have high mass loss 
rates (Mowlavi \etal 1998; Greggio \& Renzini 1990) and, consequently, 
should form lower mass central stars 
than their low metallicity counterparts.  As a result, the luminosity of the 
central star (and the nebula) should be lower if the progenitor is metal-rich.
But, as noted just above, nebular cooling is somewhat more efficient 
when there are large numbers of atoms. The two PN components,
the nebula and the central star, therefore behave in offsetting ways.
A detailed analysis shows that the peak in the [O~III] luminosity is
expected for PNe with metallicities between the LMC and solar
(Dopita, Jacoby, \& Vassiliadis 1992; Ciardullo \& Jacoby 1992).

Figure~6 shows how the apparent brightnesses of the M31 PNe vary with
O/H and Ne/H.  We include SRM objects in our plot because their sample
includes more of the brightest PNe than are present in our dataset.
(We intentionally selected targets to extend over a wide range of
luminosity in order to probe luminosity-dependent abundance patterns.)
The observations demonstrate that the brightest PNe are nearly independent
of O/H{}.  That is, the upper envelope of magnitudes is flat to within
$\pm 0.1$ mags over a range of 1 dex in O/H{}.  Thus,
we support the conclusion of Kennicutt \etal (1998), who found that the
differences between PNLF and Cepheid distances are nearly independent
of metallicity.  These data, coupled with the correlation between maximum
[O~III] flux and O/H found at lower metallicities (Richer,
Mc~Call, \& Arimoto 1997), also confirm the conclusion of Ciardullo \&
Jacoby (1992), that the sensitivity of the PNLF cutoff to metallicity
is only important for populations more metal poor than the LMC.

On the other hand, Figure~6 also demonstrates that there is a slight 
tendency for fainter PNe to be drawn from a lower metallicity population.  
If real, then the shape of the PNLF could be metallicity-dependent even if 
the maximum [O~III] luminosities remain insensitive to abundance.
We suspect, though, that this trend is an artifact of a selection effect
or an analysis bias because there is no physical rationale why metal-rich
PNe should not be faint as well as bright.  Most likely, an unbiased
distribution in O/H drops off steeply towards high O/H (as it does in our
Galaxy).  Because the SRM sample is larger than ours, it will include more
high O/H PNe than ours, and since it also draws from a brighter sample
than ours, an apparent deficit of faint, high O/H PNe may be the result.

\subsubsection{luminosity and central star mass}

Figure~7 illustrates the absolute magnitude, $M_{5007}$, of the PNe as a
function of central star mass, corrected for extinction and distance. We
see a clear trend of the sort one expects -- that is, more massive central
stars produce more luminous planetaries. At face value, this suggests that 
young populations should produce luminous PNe (at $\lambda 5007$) which 
would make the cutoff in the planetary nebula luminosity function depend
on population age.  Yet, when we look at Figure~8, which shows the apparent PN
magnitudes, $m_{5007}$, as a function of central star mass, the trend is much 
less evident.  This suggests that the PNLF has, at best, only a mild
dependence on stellar age.  Quantitatively speaking, if we ignore the
stars with core masses $>$0.8 M$_\odot$, then there is a 99.93\% probability
that $M_{5007}$ is correlated with core mass (i.e., a $3.4 \, \sigma$ effect)
but only a 92.72\% probability that $m_{5007}$ is correlated (a $1.8 \, \sigma$
confidence level).  This behavior can be understood by examining the
only parameter that can effect a difference between Figures 7 and 8 --
extinction.

\subsubsection{extinction and central star mass}

Figure~9 shows that a correlation exists between extinction to the individual
M31 PNe and their central star masses.  It operates in exactly the manner
expected -- lower core mass objects have less extinction. Thus, the lower
central star luminosities of low core mass PNe are partially compensated
by also having less extinction to dim their nebulae.  A corollary to this
observation is that the luminosities will be less constant at other
wavelengths since extinction is highly dependent on wavelength.
For example, one might expect that, when observed in C~III] $\lambda 1909$,
the brightest planetary nebulae will be those with low-mass central stars,
since the impact of extinction is greater in the UV.

There is some circularity in the relationship between extinction and
central star mass. A positive error in the Balmer decrement can induce
an overestimate of the extinction.  This error will, in turn, 
cause the extinction-corrected nebular luminosity to be overestimated, 
and hence, the luminosity of the central star will be overestimated.  Since
core mass is derived from a star's position in the HR diagram, an
overestimate of central star luminosity leads directly to an
overestimate of central star mass.  Consequently, errors in 
the data will introduce a correlation of the type seen here, but at a
level about 4--6 times smaller than observed.

Ciardullo \& Jacoby (1999) discuss this correlation in detail.  A plausible
rationale for its existence can be described briefly in the following manner.
Large core mass PNe are presumed to derive from relatively massive progenitors.
Most of the progenitor's mass is ejected during the AGB phase; consequently,
PNe with slightly higher mass central stars have considerably more material
in and around their nebulae.  Moreover, the timescale for an AGB star to
evolve into a planetary nebula is extremely sensitive to mass.
Low mass stars evolve to the blue very slowly, giving their circumstellar
matter time to disperse, while high mass stars go from the AGB phase to
a high-excitation PN quickly, when their ejecta is still close by.  As a 
result, PNe formed from high mass stars will be surrounded by a large, dense
dust cloud, while low mass PNe will have less circumstellar dust, and
have it at lower density.  A correlation between extinction and core-mass 
is the natural result.

\subsection{Abundance summary}

The average (by number, not log) and median abundances of nitrogen,
oxygen, neon, argon, sulfur, and helium are given in Table~3.  (Recall
that the values for carbon are not meaningful, except to document the
abundances used in the models.  That is, they are assumed, rather than
calculated.)  For comparison, Table~3 also includes the abundance values
for the Sun, and average values for PNe in the Milky Way, the LMC, the
SMC, the Sagittarius dwarf, and NGC 5128 (Cen A).  Also listed are the
abundances of the Orion H~II region. 

The M31 helium abundances are comparable to, or slightly higher than,
those in the Milky Way PNe.  On the other hand, the median abundances
for the remaining elements are in surprisingly good agreement with the
average for the LMC sample given by Clegg (1992). The recent re-analysis
of the LMC and SMC PNe by SRM, though, shows 2 significant effects
in this regard. First, SRM find that luminous PNe are more metal-rich
than low luminosity PNe (see also \S 5.1.2), which suggests that a direct
comparison of abundances is risky unless one compares PNe of similar
luminosity.  Second, they find that the PNe in their high-luminosity sample
have O/H 0.15 dex smaller and N/H $\sim 0.35$ dex smaller than the values
reported by Clegg.  Since the SRM high luminosity sample should be more 
metal-rich than the Clegg PNe, this indicates that analysis methods can play 
an important role in these comparisons.
(We adopted Clegg's values for Table 3 because he presented a single
set of abundances that includes neon, sulfur, and argon, whereas SRM
present only oxygen and nitrogen).

The similarity of the M31 PN abundances to those of LMC PN is unexpected
in light of the high abundances derived from the integrated stellar
spectra. We discuss this conundrum below.

\section{Discussion}

\subsection{LMC abundances for a super metal-rich population?}

The abundances derived from the integrated light of galaxies
generally yield calibrated abundances only for [Fe/H]. While absorption
indices such as CN and CO provide limited information about the lighter
elements (Burstein 1985), a direct quantitative comparison for light
element abundances between the stars and PNe is not possible.
Consequently, we will compare [Fe/H] from the stars in M31 to 
[O/H] from the PNe.

The primary question raised by Table~3 is the following: given that
the integrated properties of the bulge starlight in M31 exhibit clear
signatures of high metallicity, with [Fe/H] $\sim +0.25$ (see \S 1),
why do the PNe in M31 appear to be so similar to the LMC PNe{}? After
all, the LMC is generally considered to be dominated by stars with
[Fe/H] $\sim -0.4$ (Olszewski, Suntzeff, \& Mateo 1996) and surveys
in this galaxy demonstrate that the PNe have similarly low abundances
(Dopita \& Meatheringham 1991a; Dopita \& Meatheringham 1991b; Clegg
1992; Dopita \etal 1997).  The same can be said for the Galactic bulge:
the mean metallicity of the stars is [Fe/H] $\sim -0.25$ (McWilliam \&
Rich 1994) while the Galactic bulge PNe have [O/H] $\sim -0.3$ (Ratag
\etal 1997).  But in M31, there is a clear misalignment of $>$0.5 dex
between the expected and observed abundance patterns.  We consider
several explanations for this discordance.

\subsubsection{Metal-rich stars cannot make bright PNe}

We must keep in mind that we are only surveying the brightest PNe in
M31's bulge. Thus, the entire issue can be side-stepped if metal-rich
stars do not make bright PNe. Although far from certain, if mass loss
prior to the PN phase increases with metallicity (Mowlavi \etal 1998;
Greggio \& Renzini 1990), then at some high level of metallicity, only
very low mass remnant stars will be produced. These will result in faint
or possibly ``invisible'' PNe.  Moreover, if the central star falls below
a critical mass ($\sim 0.55$ M$_\odot$), it may become a ``lazy central
star'' (Sch\"onberner 1983), which will evolve too slowly to ionize its
ejecta.  This effect is likely to occur in an old, low mass population
(Jacoby \etal 1997).  Thus, it is possible that a large fraction of
M31's high-metallicity bulge stars won't produce bright planetaries.

There is supporting evidence for this scenario -- CJFN estimated that
the PN production rate of M31's bulge is about half that expected from
the total stellar death rate (Ciardullo 1995).
Similarly, despite the large range of metallicity observed
for Milky Way bulge stars, the survey by Ratag \etal (1997) found only one
PN out of 103 where O/H exceeds solar by more than the error bar.

Figure 10 shows the distribution function for O/H in the M31 and Milky
Way bulges.  The total M31 sample (ours plus SRM's) is statistically
identical to that of the Milky Way: a Kolmogorov-Smirnov test shows
that the two distributions are different only at the 42\% confidence
level (0.55 $\sigma$) -- that is, they cannot be distinguished with any
confidence.  Moreover, both distributions contain very few super-solar
PNe, especially when one considers that the uncertainty in the 
oxygen abundances is typically 0.1 dex or greater for the M31 sample
and $\sim 0.3$ dex for the Galactic bulge sample.

Another comparison of PN and stellar abundances measurements can be
attempted in M32.  Freedman's (1989) distribution of stellar metallicity,
[M/H], shown in her Figure~8, is similar to the PN O/H distribution seen
here in Figure~10.  Based on various color indicators, Freedman derives
a mean metallicity for M32 of [M/H] $\sim -0.5$.   This compares to
the [O/H] value of $-0.66$ obtained by SRM from spectroscopy of 9~PNe.
This difference becomes more significant when one considers that solar
neighborhood stars with [M/H] $\sim -0.5$ have [O/H] $\sim -0.4$,
and suggests that the differential in metallicity indicators is $-0.26$
dex in M32 relative to the Galaxy.  Moreover, the colors obtained by
Grillmair \etal (1996) for an inner field of M32 ($1-2\arcmin$ from the
nucleus) indicate that that galaxy's [M/H] abundance is significantly 
higher ($-0.25$), thus implying an even stronger disagreement with the PN 
[O/H] measurement.

Therefore, for M31's bulge and possibly M32 as well, the PN abundances appear
to be $\sim 0.3$ dex lower than the mean metallicity of the stars.  This
seems to indicate that either metal-rich stars do not produce bright PNe or
there is a zero-point offset in the two different approaches to deriving 
metallicities.

\subsubsection{There is a selection bias against metal-rich PNe}

We have already noted that our sample of PNe was chosen to cover a
relatively wide range of [O~III] $\lambda 5007$ luminosity, more-or-less
uniformly.  Consequently, the 15 PNe studied in this survey extend more 
than 2~mags across the PNLF, and nearly 3~mag below the maximum 
[O~III] $\lambda 5007$ brightness attainable by a planetary.  This
contrasts with the SRM sample, which is concentrated toward the
top magnitude of the PNLF{}.  On average, the SRM sample of PNe have higher
abundances than the PNe we observed, with a median O/H of $\sim 8.7$ and median 
Ne/H of $\sim 7.9$ (see Figure~6).  In other words, the SRM sample is
$\sim 0.3$~dex more metal-rich than our own; this suggests that interpreting 
our results as ``LMC-like'' may be overly simplistic.

The most fundamental conclusion that we can safely state is that there
is a considerable spread in the abundance properties in the M31 bulge PNe.
Since elements such as oxygen, neon, and argon are not changed significantly
within their parent stars prior to ejection of the nebula (van den Hoek \& Groenwegen 1997; Forestini \& Charbonnel 1997; Aller 1994),
a large spread in progenitor abundances is likely.
Thus, the stars in the bulge of M31 must have a wide range of
properties, and that referring to the ``metallicity of the bulge'' as
a single parameter makes no sense.  One must consider the stellar
populations problem more generally and refer to each elemental abundance 
as a distribution function.

Given that there is a spread in the stellar abundances of M31's bulge, it
is possible that selection effects operate to favor the observation of
low-to-moderate metallicity objects. In fact, one such selection effect
has already been discussed; in \S 4.4.1, we noted that the most favored
abundances for bright PNe occurs between the LMC and solar metallicity.

\subsubsection{[O/H] and [Ne/H] don't correlate directly with [Fe/H]}

For Galactic field stars, [O/Fe] decreases with increasing [Fe/H] (Edvardsson
\etal 1993; King 1994; Gratton \etal 1996). At very low metallicities,
[O/Fe] $\sim +0.5$. When [Fe/H] $\sim +0.2$, though, [O/Fe] has dropped to
$\sim -0.2$, giving [O/H] $\sim 0.0$.  At higher metallicities, one presumes
that the trend continues, but data at [Fe/H] $>+0.2$ is very sparse.
If we assume that the bulge of M31 has a similar chemical enrichment history
as the nearby field stars, then we should expect to see M31 bulge PNe with
[O/H] $\sim 0.0$ (i.e., O/H $\sim 8.9$). In fact, we do see several
objects in this regime, and perhaps these oxygen-rich PNe derive from 
iron-rich progenitors while the oxygen-poor PNe truly derive from
iron-poor progenitors.  Unfortunately, we do not have iron abundances for 
the PNe, nor do we have oxygen abundances for the stars.

On the other hand, if the PNe studied in our survey come from stars
with high [Fe/H], then we are compelled to consider chemical enrichment
scenarios that differ from that of the local neighborhood. For example,
the bulge stars of M31 could have been created in an environment where the
SN~II population (a main source of oxygen) was quickly truncated relative
to the SN~Ia population (a main source of iron).  For example, if the bulge 
of M31 was formed from the accretion of many dwarf galaxies, whose early 
history was dominated by supernova driven winds, then it is possible that
star formation could have been terminated before oxygen was fully 
enriched.  In such a scenario, iron enrichment would come later, as the 
result of Type~Ia supernovae, and an anomalously low O/Fe would be the result. 
This model fails in detail, since it predicts a bulge iron abundance
similar to that seen in dwarf galaxies (\eg [Fe/H] $< -1$; Caldwell 
\etal 1998), but it does demonstrate how O/Fe might be modified.

WFG considered several solutions to the related, but opposite problem:
that of enhanced [Mg/Fe] in the nuclei of galaxies.  Since our observations
are of M31's bulge, the same mechanisms may operate to increase
Mg/Fe in the center of a galaxy and decrease O/Fe further out.
For example, the two most plausible processes described in WFG are

\begin{enumerate}

\item {a gradient in the distribution of stars with $M > 15 M_\odot$.
If few massive stars formed in the bulge of M31, then the descendent 
Type~II SNe would have produced few light elements in the bulge.  
Conversely, if massive stars were common near the nucleus, then
Mg/Fe in this region could have been enhanced.}  

\item {a slower formation time scale for the bulge compared to the
nucleus.  If the bulge of M31 was built slowly in hierarchical fashion,
then there is more time available for Type~Ia SNe to enhance iron without
drastically affecting oxygen.}
 
\end{enumerate}

Thus, the different results in the two regions could be ascribed to the
differing environments; the nucleus is small (enrichment time scales
are short) and has a high density (high mass stars may form more easily).

\subsubsection{A gradient in [O/H] in the bulge of M31?}

Historically, the high values (+0.25) of [Fe/H] in M31 are based
on observations taken very close to the nucleus, and may even represent
a nuclear population that is very different from the surrounding bulge
stars.  Our PNe
are typically $\sim 150''$ from the nucleus.  Davidge (1997) measured
the gradients in [Fe/H] for the bulge of M31 out to $60''$, and if
we extrapolate his results to a radial distance of $150''$, then we
predict that [Fe/H]$\sim 0.0$. Considering that the PNe also have a line
of sight distance from the nucleus that we are ignoring, then a solar
composition might be the largest we should expect.  Furthermore, King
(1994) shows that stars in the solar neighborhood have [O/Fe]$\sim -0.1$
when [Fe/H]$\sim 0.0$. In the presence of an abundance gradient in M31,
we may need to redefine our expectations for the PNe to a 
median O/H$\sim 8.8$, or smaller.

Further support for a significant abundance gradient comes from Mould \&
Kristian (1986), Reitzel, Guhathakurta, \& Gould (1998), and Holland,
Fahlman, \& Richer (1996) who found that stars in the outer spheroid ($>7$
kpc) of M31 have [Fe/H]$\sim -0.6$ and that the spread in metallicity
is nearly 2 dex. Thus, between the very center of the galaxy and $\sim7$
kpc out, the metallicity has dropped off rapidly. The median [O/H] value
derived from the PNe is $-0.5$ to $-0.2$, depending on the PN sample. At
a typical galactocentric distance of 0.5 kpc, this [O/H] is intermediate
between the metal-rich and low metallicity stellar populations.

\subsection{The halo PNe}

Among our 15 PNe, there appear to be two candidates for classification
as halo objects.  This is somewhat surprising in that, in our Galaxy, 
less than 1\% of the known PNe belong to this group.  The most curious
halo candidate is CJFN 478, having abundances that are 4 to 18 times lower
than solar. The central star mass, however, is deduced to be
$ 0.88 M_\odot$, and if we adopt the initial mass-final mass relation of
Weidemann \& Koester (1983), the progenitor mass is derived to be 
$M \sim 5 M_\odot$.    Clearly, this value is inconsistent with
membership in an old halo population. 

While the properties of this bulge PN could be explained by invoking
a recent star formation episode or capture of a young, low-metallicity
satellite system, neither of these ideas has strong supporting evidence.
CJFN 478 does have one of the highest velocities among our 15 PNe, with $v
= +15$~\kms\ observed, or +305~\kms\ relative to the systemic velocity
of the galaxy.  This is almost three times the stellar
velocity dispersion in the region (McElroy 1983). Given 12 bulge PNe,
one is unlikely to encounter any object having a velocity $>3\sigma$
from the mean, since the probability of finding one is 0.3\%. Thus,
statistically, there is some kinematic evidence to believe that this PN
might have come from an accretion event.  Alternatively, this could be
a halo PN with a massive central star formed through a binary coalescence.

The other halo candidate is FJCHP 57, which has abundances 2 -- 4 times
below solar, except for neon which is highly deficient ($40\times$
below solar).  As noted in \S 4.3, FJCHP 57 is peculiar in that it
lies well off the oxygen--neon correlation. Furthermore, FJCHP 57 has
a velocity $\sim 400$~\kms\ more positive than the other two disk PNe,
which are within 70~\kms\ of each other. Thus, it is likely that this
object comes from a different kinematic population.

Given that we have stumbled on 2 halo PNe out of a sample of 15, one
might suspect that the halo component of M31 is very luminous compared
to that of the Galaxy. It is premature, though, to comment further on
the halo population of M31 until a larger sample of PNe are observed,
both for abundances and kinematics.

\subsection{Disk vs bulge abundances}

Interestingly, if we discount the halo object FJCHP 57, then the two
disk PNe, which are at a projected galactocentric radius of 14~kpc,
have essentially the same median abundances as the 12 bulge PNe. That
is, with the limited sample from this paper, there is no evidence for an
abundance gradient between 0.5 kpc from the center of M31, and a region
well beyond where the solar circle would fall.

On the other hand, if we mix the SRM PN sample in with ours, then there
is a mild gradient in the medians. Since the distribution of abundances
in the bulge extends both well above and well below the abundances of
the two disk PNe, it is clear that a simple gradient in the median fails
to characterize the two samples properly.  One needs dozens of PNe, both
in the bulge, and at a variety of galactocentric radii, to define
the chemical gradient of M31's stars.  Given the number of PNe available
for study, such measurements are feasible.

A previous attempt to measure the abundance gradient in M31 from PNe (Jacoby
\& Ford 1986) also found a small gradient, but that was based on only
three PNe. Other methods (HII regions, supernova remnants) identify a
mild gradient as well (Dennefeld \& Kunth 1981; Blair, Kirshner, \& 
Chevalier 1982).

\subsection{Evidence for a merger event?}

Our limited sample of 13 bulge PNe shows no evidence for spatially
correlated abundances. If we include the additional 25 bulge PNe from
SRM (their 28 bulge PNe include 3 of ours), one abundance coincidence
emerges. Within a 1 arcmin region (220 pc) containing the PNe CJFN 28,
29, 30, 31, and 80, the average O/H is $8.95 \pm 0.11$; for comparison,
the surrounding PNe typically have O/H $\sim 8.35$. Thus, over a 10
arcmin field from which the sample of 38 PNe were selected, half of
the PNe with O/H $> 8.80$ are found within 1 arcmin of each other; the
remaining high metallicity PNe are distributed fairly randomly across the
bulge field (although mostly to the west of the nucleus). Velocities of
the 5 PNe comprising the high O/H concentration will provide a critical
test of the possibility that these PN are remnants of a merger event.
If the kinematics confirm an association among these 5 PNe, then we know
from the PN production rate that the accreted object is comparable in
luminosity to the M31 dwarf elliptical companions NGC 147 and NGC 185
(CJFN; Ford, Jacoby, \& Jenner 1977).

On the other hand, the metallicity--luminosity relationship among
galaxies argues that a dwarf elliptical galaxy would not have the high
abundance levels exhibited by these 5 PNe. The merger prospect can still
be salvaged if the PN grouping derives from the central dense region of
a larger galaxy that otherwise has been disrupted and dispersed into
the general bulge population. Thus, the possible merger of metal-rich
stars is far from settled.

\subsection{Future directions}

The M31 study presented here, plus that published by SRM, illustrate
that PNe provide a direct way to probe elemental abundances, their
distributions, and their gradients in old populations. Furthermore Walsh
\etal (1998) derived abundances for 5 PNe in the nearby peculiar elliptical
galaxy, NGC 5128 (Cen A), demonstrating that 4-m class telescopes can
be used to investigate the properties of individual stars in the halos
of galaxies four times further than M31. With 8-m class telescopes at excellent
sites, one can expect to derive the properties of the old stellar
populations in many more elliptical and spiral galaxies.  Such observations
should completely clarify the ambiguity about the color gradients
of galaxies, \ie whether they are due to age differences
or compositional differences.

\section{Conclusions}

The principal conclusions of this paper are:

\begin{enumerate}

\item The very large dispersion in abundances for the bulge PNe of M31
indicates
that the stellar population must be highly diverse and that one cannot
characterize the bulge with a single metallicity. Instead, it is far
more appropriate to refer to the distribution functions of each element.

\item The PNe in the bulge of M31 have abundances that are typically
sub-solar.  The median abundances could be as low as those in the LMC
([O/H] $\sim -0.5$), or, if the sample is restricted to only the most
luminous objects, the abundances may be 0.3 dex higher than the LMC{}.
Several factors can account for the apparent disparity between this
value and the super-solar abundance measured for iron.  Among these is
the possibility that metal-rich stars fail to produce bright PNe, that
PNe with sub-solar abundances are preferentially found in flux-limited
samples, and that [O/Fe] drops when [Fe/H] is high.  An abundance gradient
within M31's bulge may also contribute to the discrepancy.  In any case,
the metallicity dispersion of the PNe in M31's bulge is significantly
larger than that seen in the LMC{}.

\item The maximum [O~III] $\lambda 5007$ luminosities attained by PNe 
are highly 
independent of their oxygen abundances. There is, however, an unexplained 
trend that the median abundances increase with increasing luminosity.

\item The PN luminosity increases as central star mass increases 
(\ie PNe from massive progenitors are brighter), but when apparent 
magnitudes are measured, this trend is offset by increased extinction.

\item There may be a mild abundance gradient from the bulge of M31 out to
the disk field at 14 kpc.

\item There is a clump of 5 spatially coincident high-metallicity PNe
about 2.5 arcmin northeast of the nucleus of M31. If they prove to have
similar radial velocities, these objects would indicate that a metal-rich
satellite galaxy has recently merged with the bulge of M31.

\item We tentatively identify 2 halo PNe in our sample of 15. This ratio
is much higher than seen in the Galaxy, and if correct, is either
fortuitous or indicative of a much larger halo population in M31 than
in the Galaxy.

\end{enumerate}

\acknowledgments

Discussions with Gary Ferland, Paul Harding, Rob Kennicutt, Heather
Morrison, and Michael Richer proved invaluable at various stages of
this study.  Comments provided by Grazyna Stasi\'nska and Lawrence Aller
on an earlier version of this paper helped clarify the presentation
immensely.  GHJ wishes to
thank Dr. Peter Strittmatter for generously providing all office needs
during a sabbatical stay at the University of Arizona where most of this
paper was written. RC wishes to thank Dr. Sidney Wolff for providing
office facilities at NOAO during a sabbatical stay during part of the
development of this paper. RC was partially supported by NSF grant
92-57833 and NASA grant NAG 5-3403.

%\end{document}

\clearpage

%----------Figure Captions-------------------------------------------

\begin{figure}
\plotfiddle{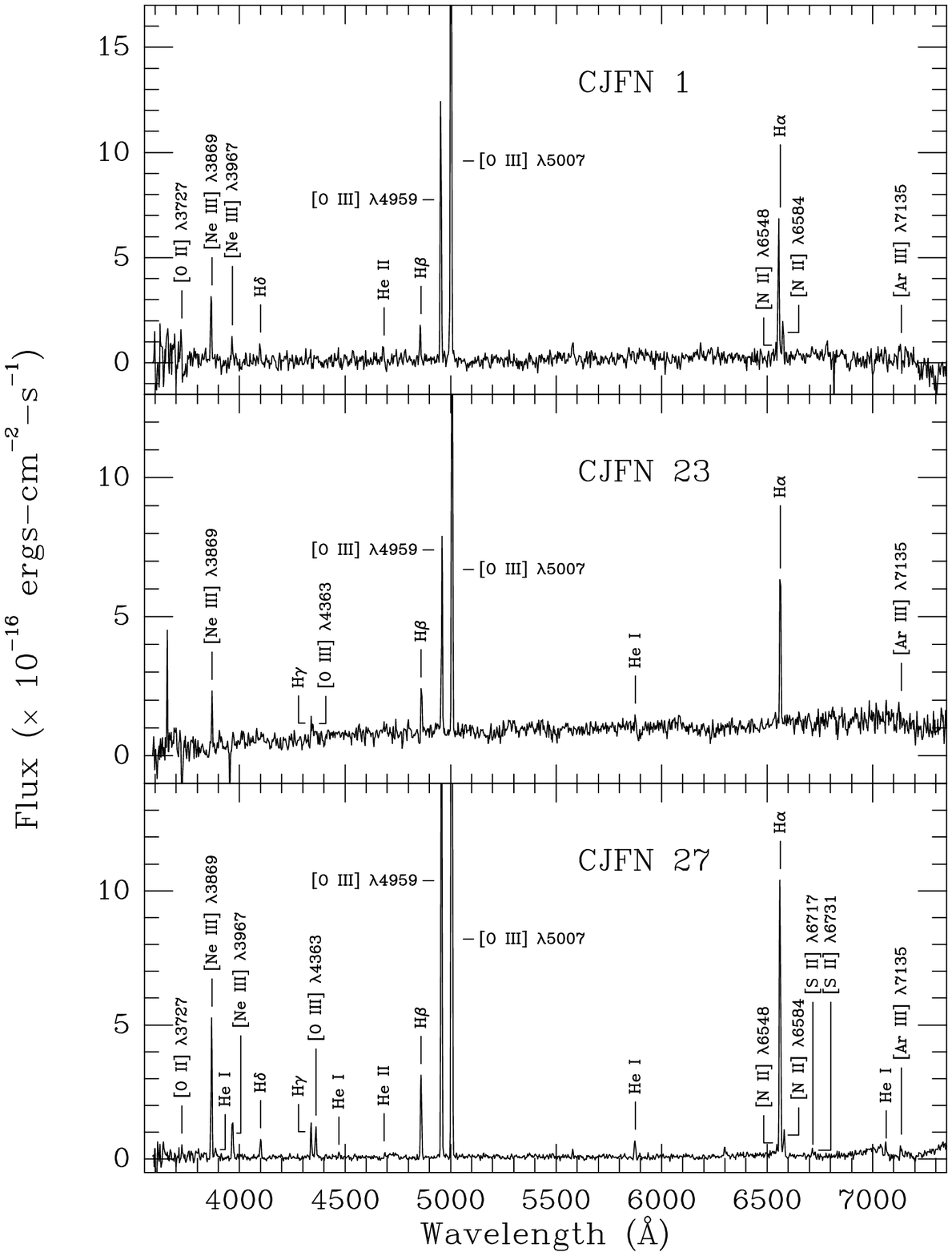}{8.0in}{0.}{95.}{95.}{-300}{-100}
\caption{Spectra of the bulge PNe, CJFN 1, 23, and 27. 
The detected lines of astrophysical interest are labeled. In some spectra, a
residual from the sky subtraction is seen at $\lambda 5577$ due to the very
strong [OI] night sky line.}
\end{figure}

\begin{figure}
\figurenum{1., cont'd}
\plotfiddle{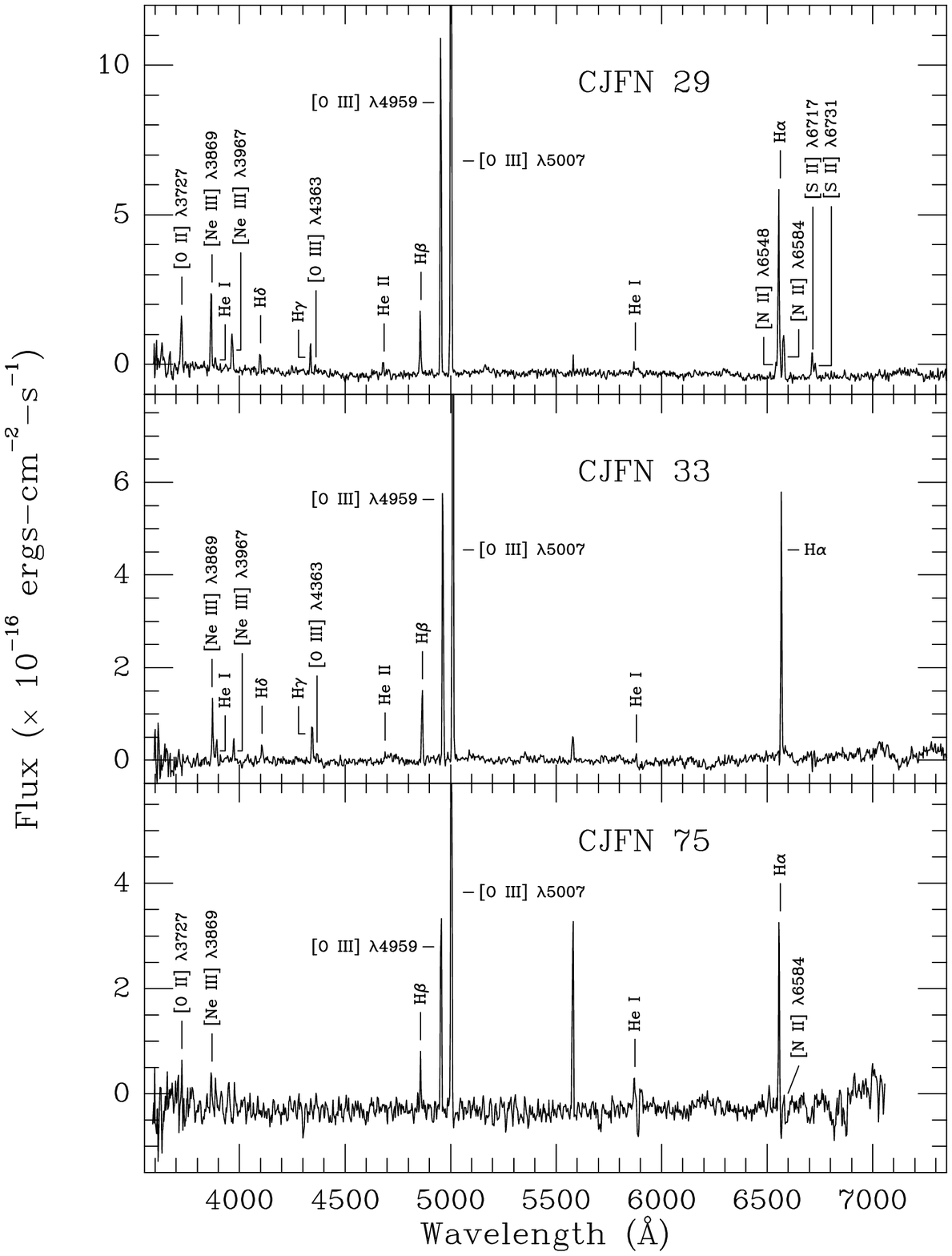}{8.0in}{0.}{95.}{95.}{-300}{-100}
\caption{Spectra of the bulge PNe, CJFN 29, 33, and 75. The [OI] $\lambda 5577$
residual from the sky is especially evident for CJFN 75.}
\end{figure}

\begin{figure}
\figurenum{1., cont'd}
\plotfiddle{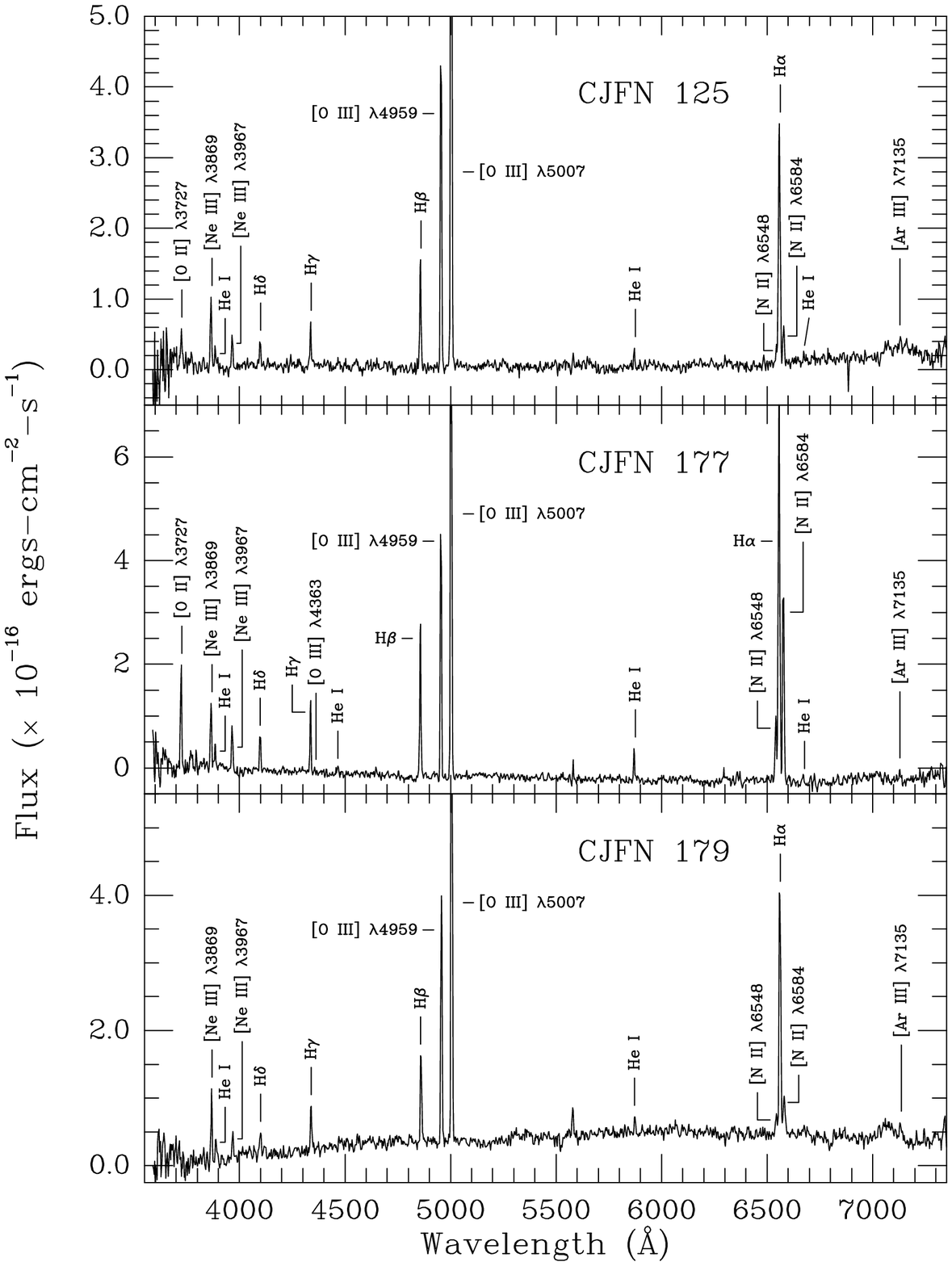}{8.0in}{0.}{95.}{95.}{-300}{-100}
\caption{Spectra of the bulge PNe, CJFN 125, 177, and 179}
\end{figure}

\begin{figure}
\figurenum{1., cont'd}
\plotfiddle{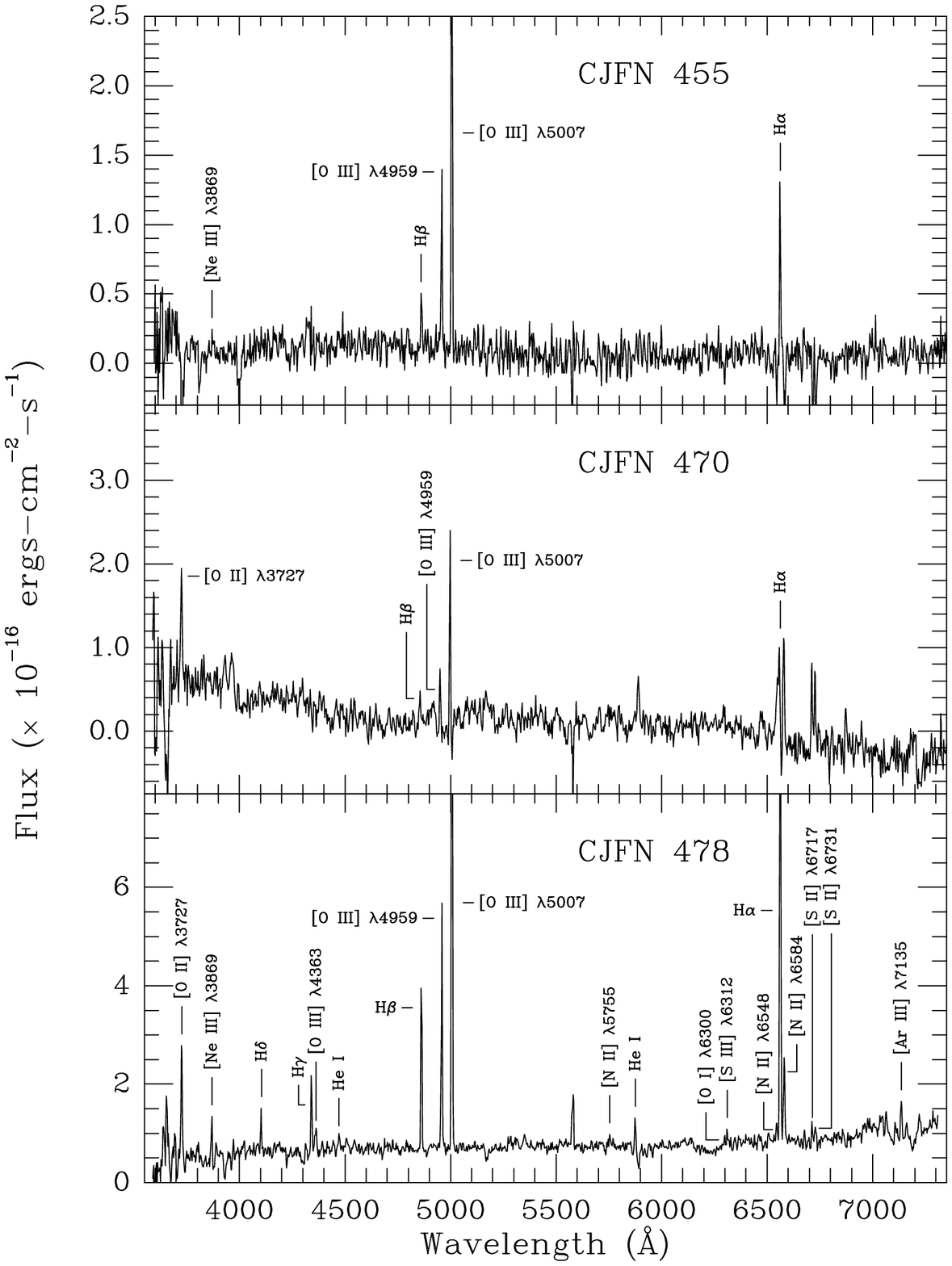}{8.0in}{0.}{95.}{95.}{-300}{-100}
\caption{Spectra of the bulge PNe, CJFN 455, 470, and 478}
\end{figure}

\begin{figure}
\plotfiddle{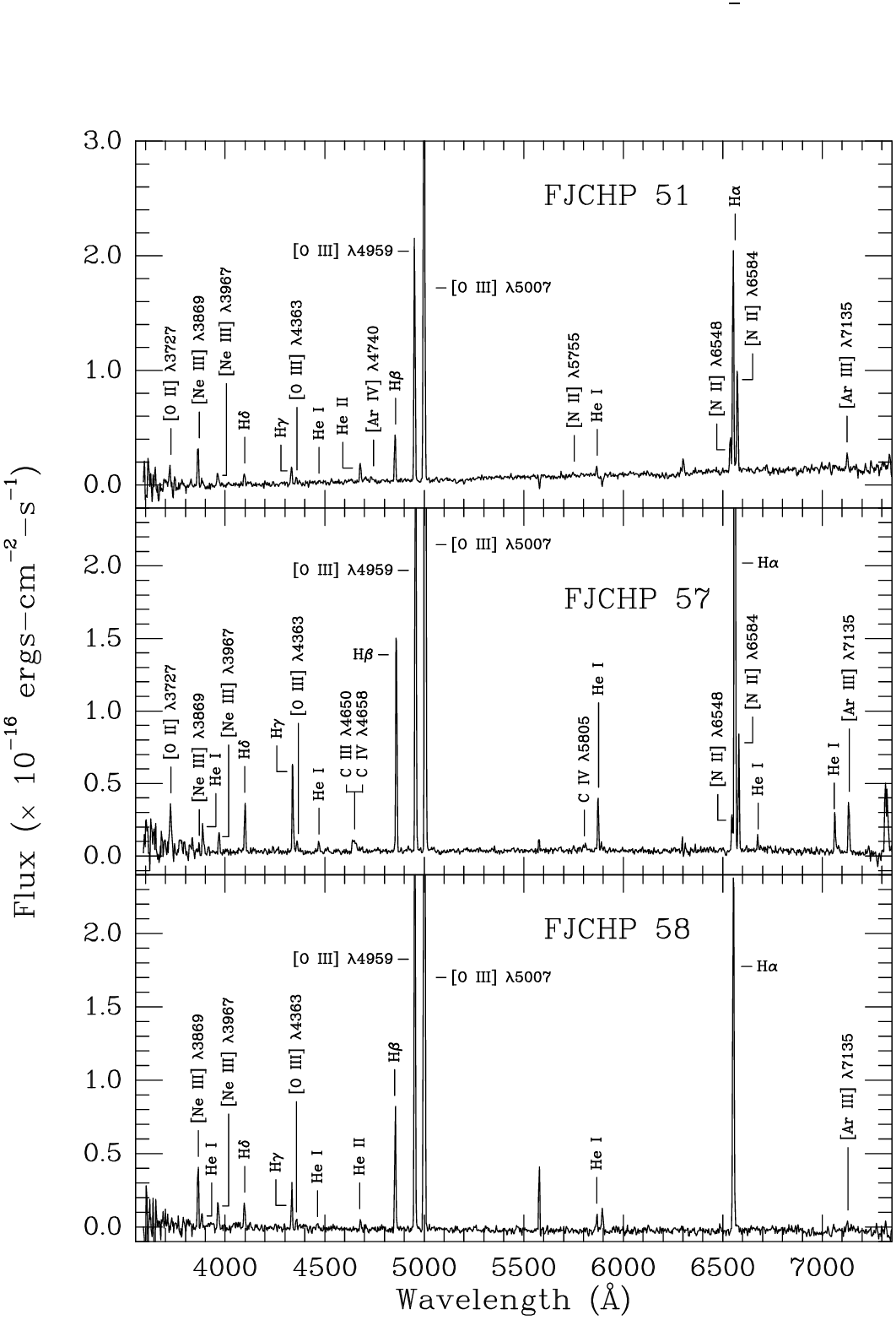}{8.0in}{0.}{95.}{95.}{-300}{-100}
\caption{Spectra of the 3 disk PNe, FJCHP 51, FJCHP 57, and FJCHP 58.} 
\end{figure}

\begin{figure}
\plotfiddle{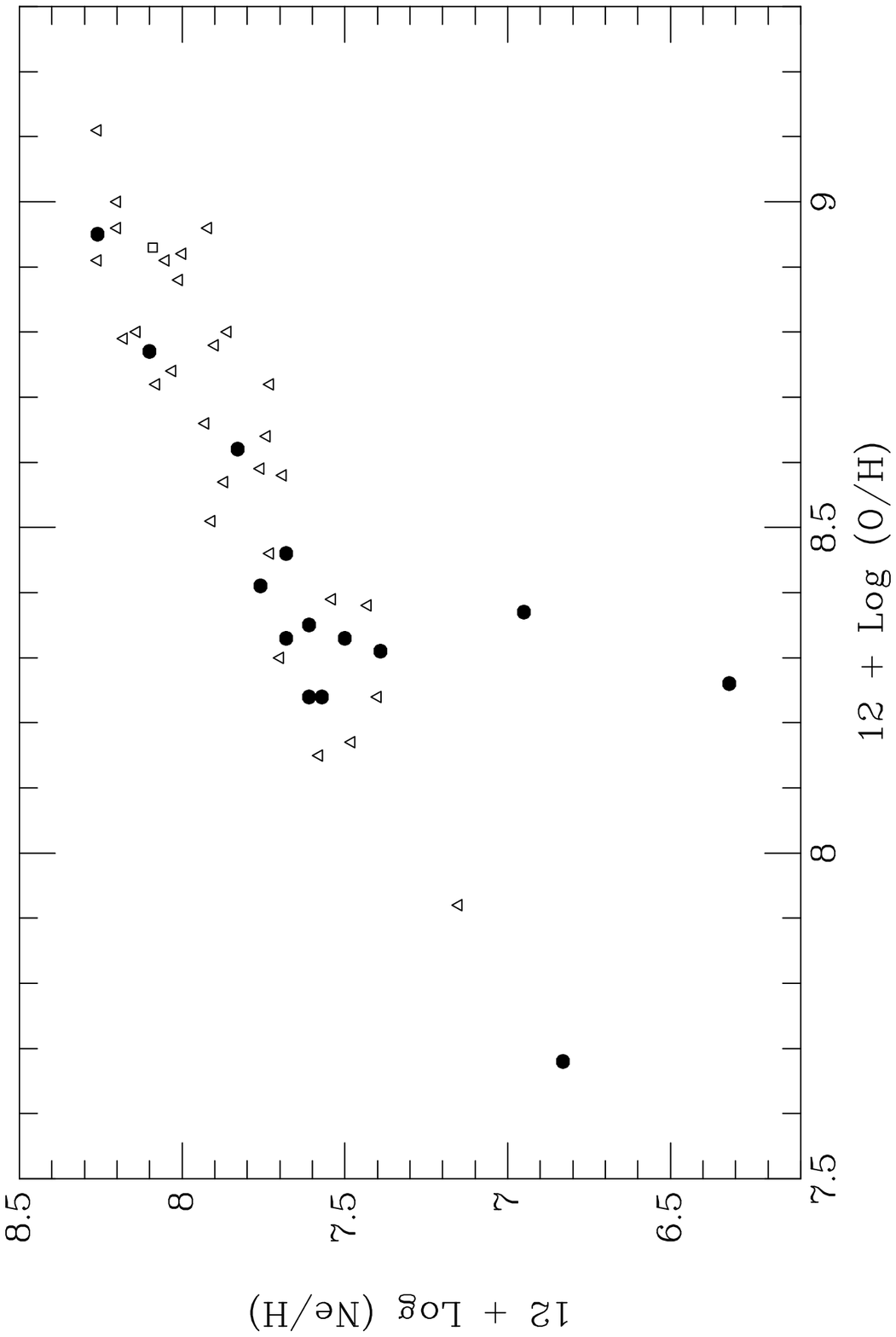}{5.0in}{-90.}{80.}{80.}{-330}{400}
\caption{The correlation between Ne/H and O/H is evident and follows
the same function as that seen by Henry (1989). The solid points represent
the PNe from this study. Open triangles are those reported by SRM for their
M31 PNe. The solid square is the Sun. The two objects that fall off the
relationship are FJCHP 57, a probable halo PN with the lowest neon abundance
in the sample, and CJFN 455, which has a marginal detection for neon.}
\end{figure}

\begin{figure}
\plotfiddle{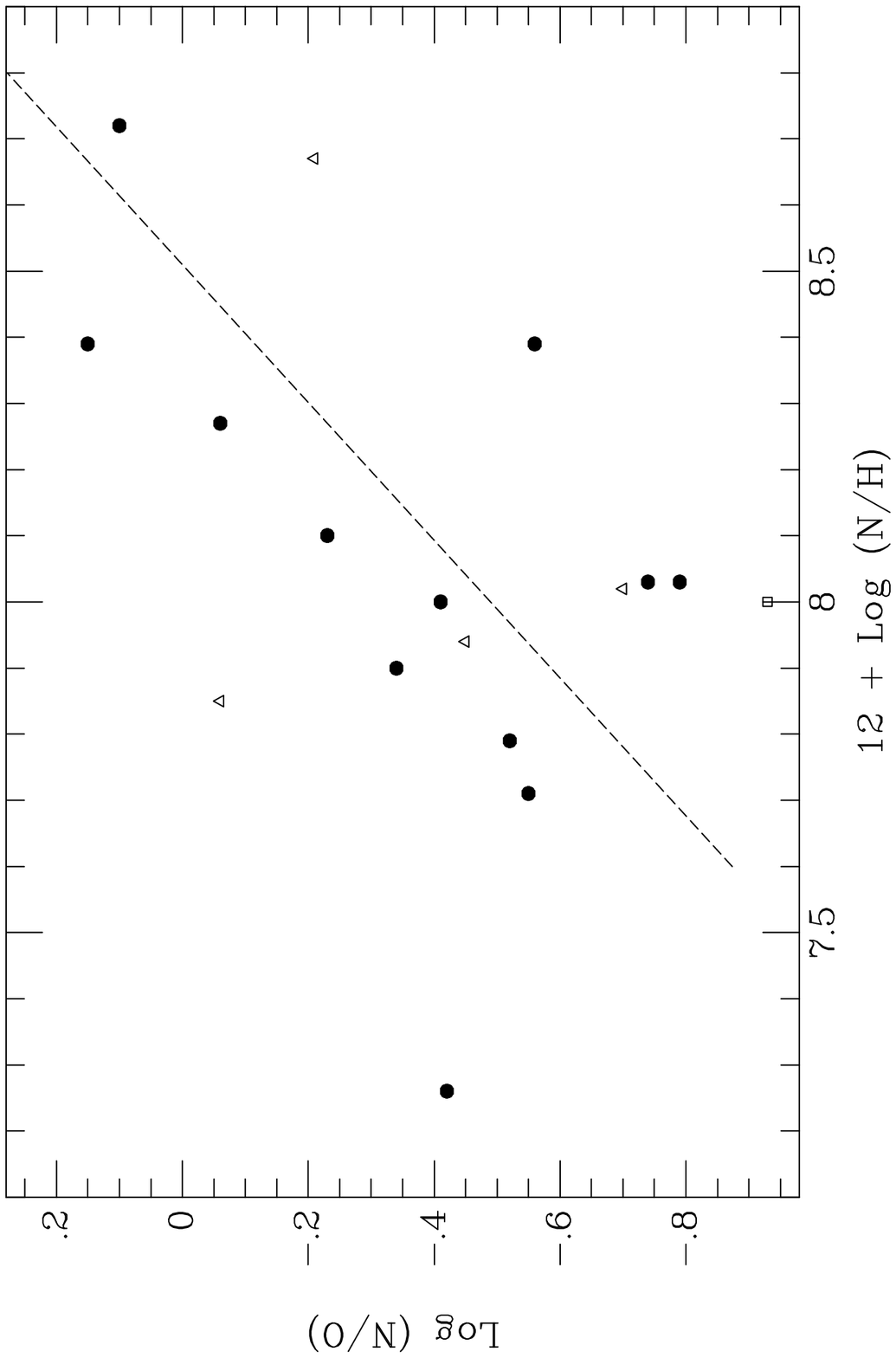}{5.0in}{-90.}{80.}{80.}{-330}{400}
\caption 
{The relationship between N/O and N/H. The dashed line shows the
fit to the Galactic data determined by Henry (1990). The point that
falls off the line to the left, with the lowest N/H, is CJFN 478, a
possible halo PN{}. Symbols are the same as in Figure~3.}
\end{figure}

\begin{figure}
\plotfiddle{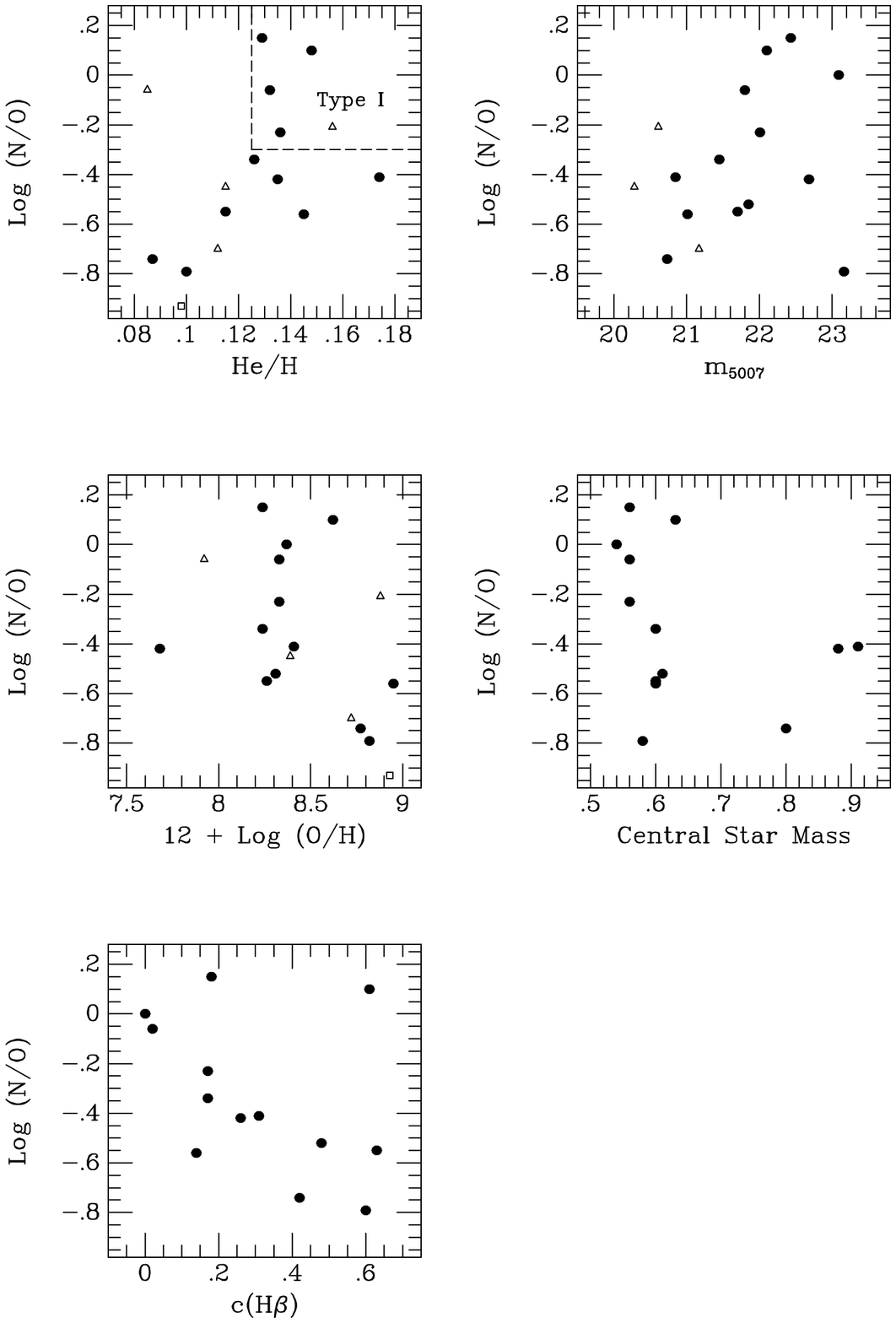}{7.5in}{0.}{90.}{90.}{-300}{-100}
\caption
{A variety of N/O relationships: (a) N/O vs He/H with the Type I regime
as defined by Peimbert \& Torres-Peimbert (1983) indicated; (b) N/O vs O/H
showing a mild, if any, anti-correlation; (c) N/O vs the extinction parameter, 
$c$(H$\beta$); (d) N/O vs $m_{5007}$, and (e) N/O vs central star mass. 
Symbols are the same as in Figure~3.}
\end{figure}

\begin{figure}
\plotfiddle{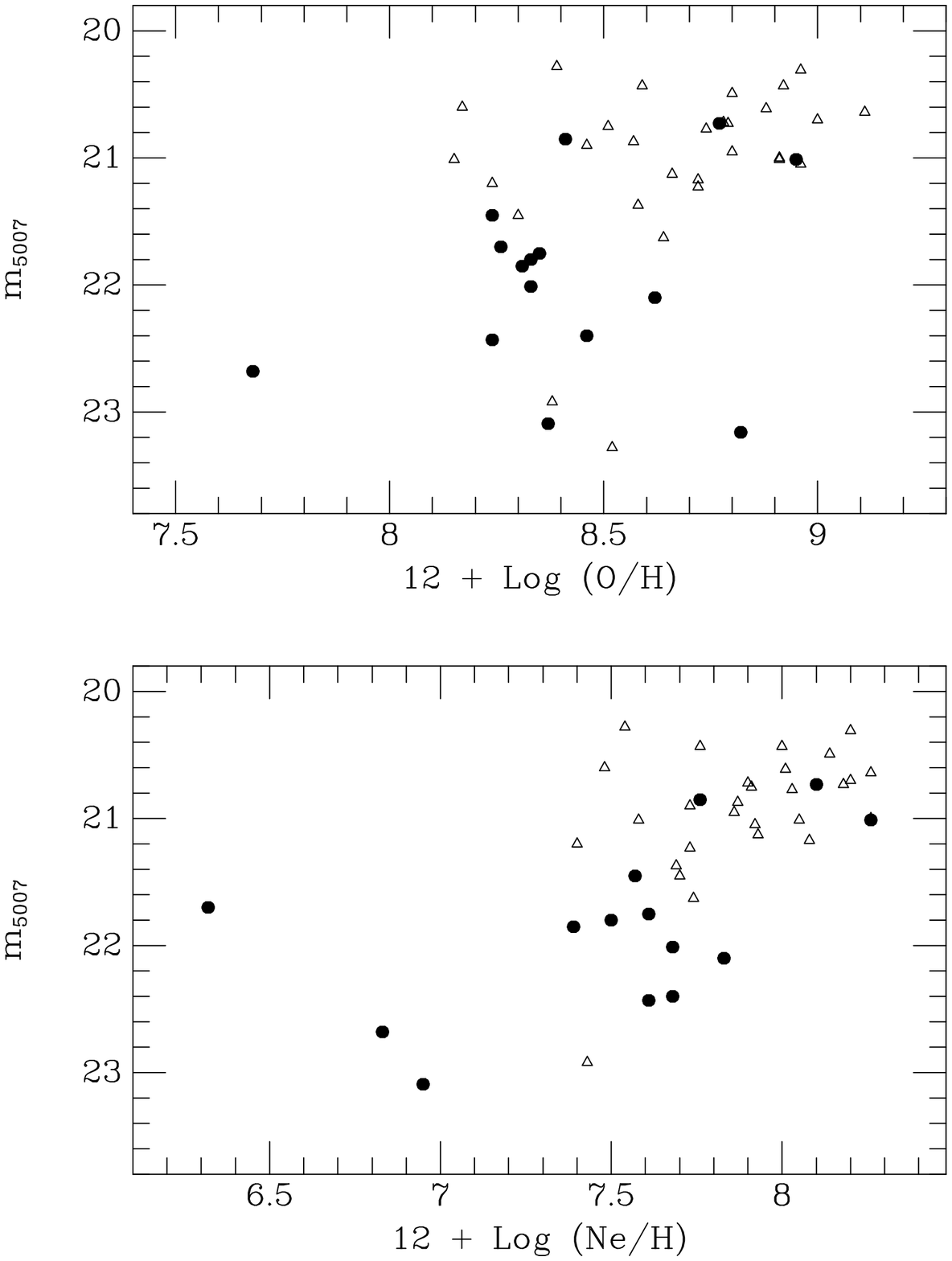}{7.0in}{0.}{85.}{85.}{-280}{-80}
\caption
{The relationships between apparent PN brightness, $m_{5007}$,
and the oxygen and neon abundances. The upper envelopes are nearly flat
as abundance changes, indicating near-independence of PN maximum luminosity
with metallicity. Symbols are the same as in Figure~3.}
\end{figure}

\begin{figure}
\plotfiddle{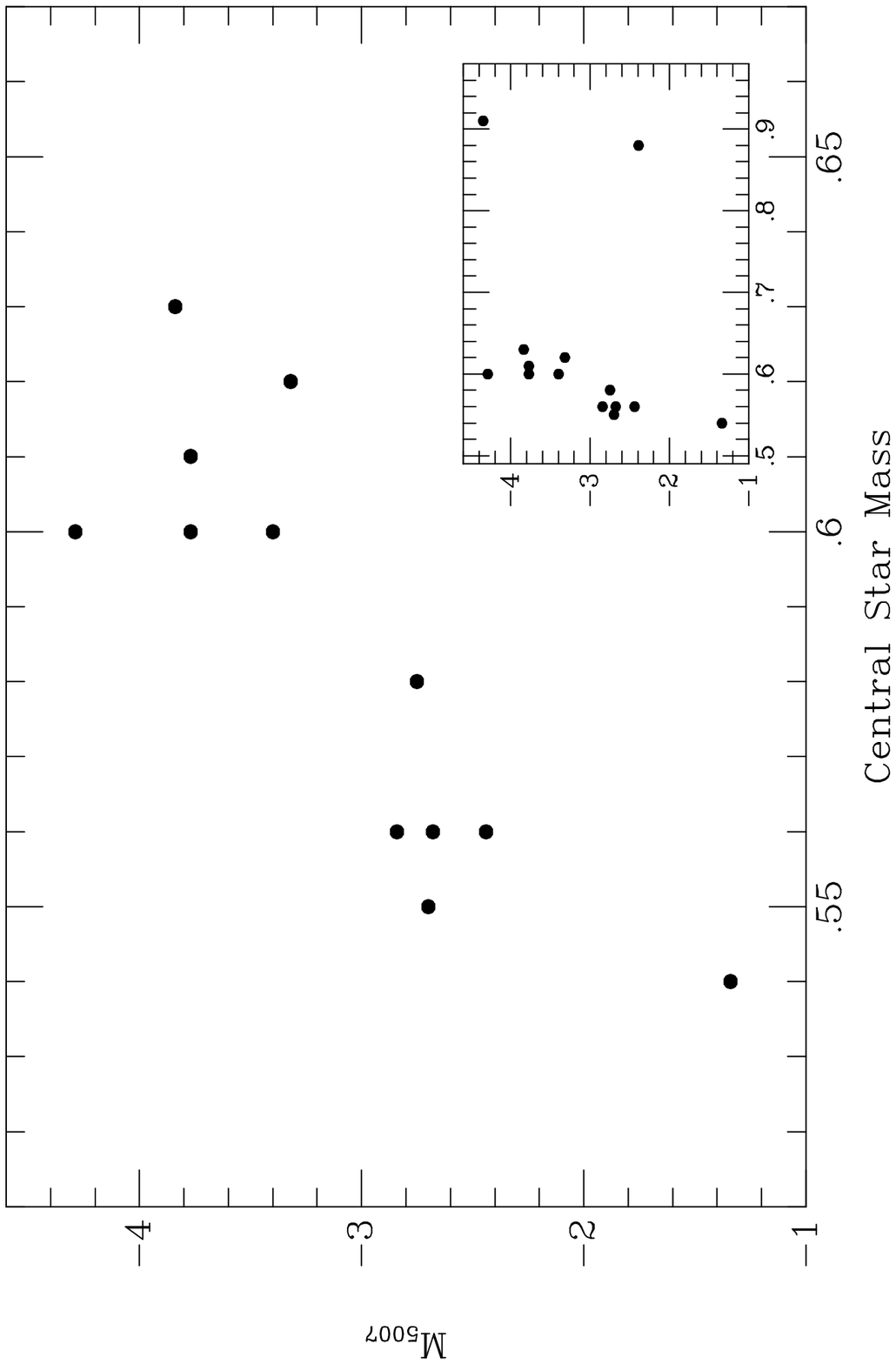}{5.0in}{-90.}{80.}{80.}{-330}{400}
\caption
{The absolute magnitude at [OIII] $\lambda 5007$ (corrected
for the measured extinction) correlates with central star mass, at least
for the low-to-moderate mass stars. The correlation may break down at higher
masses (see inset).}
\end{figure}

\begin{figure}
\plotfiddle{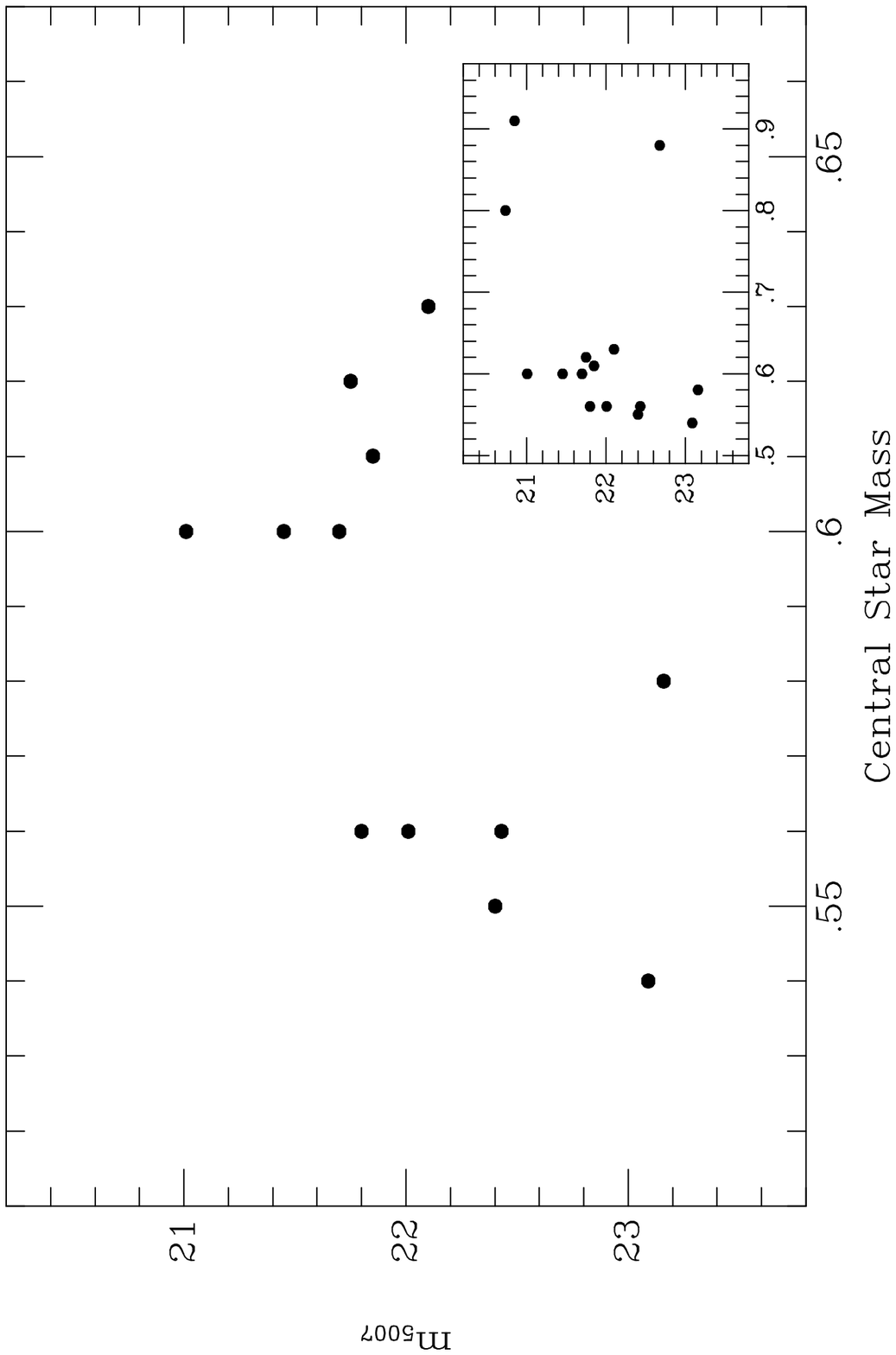}{5.0in}{-90.}{80.}{80.}{-330}{400}
\caption
{The apparent magnitude at [OIII] $\lambda 5007$ correlates, though
poorly, with central star mass. Again the high mass regime (see inset) shows
even less correlation.}
\end{figure}

\begin{figure}
\plotfiddle{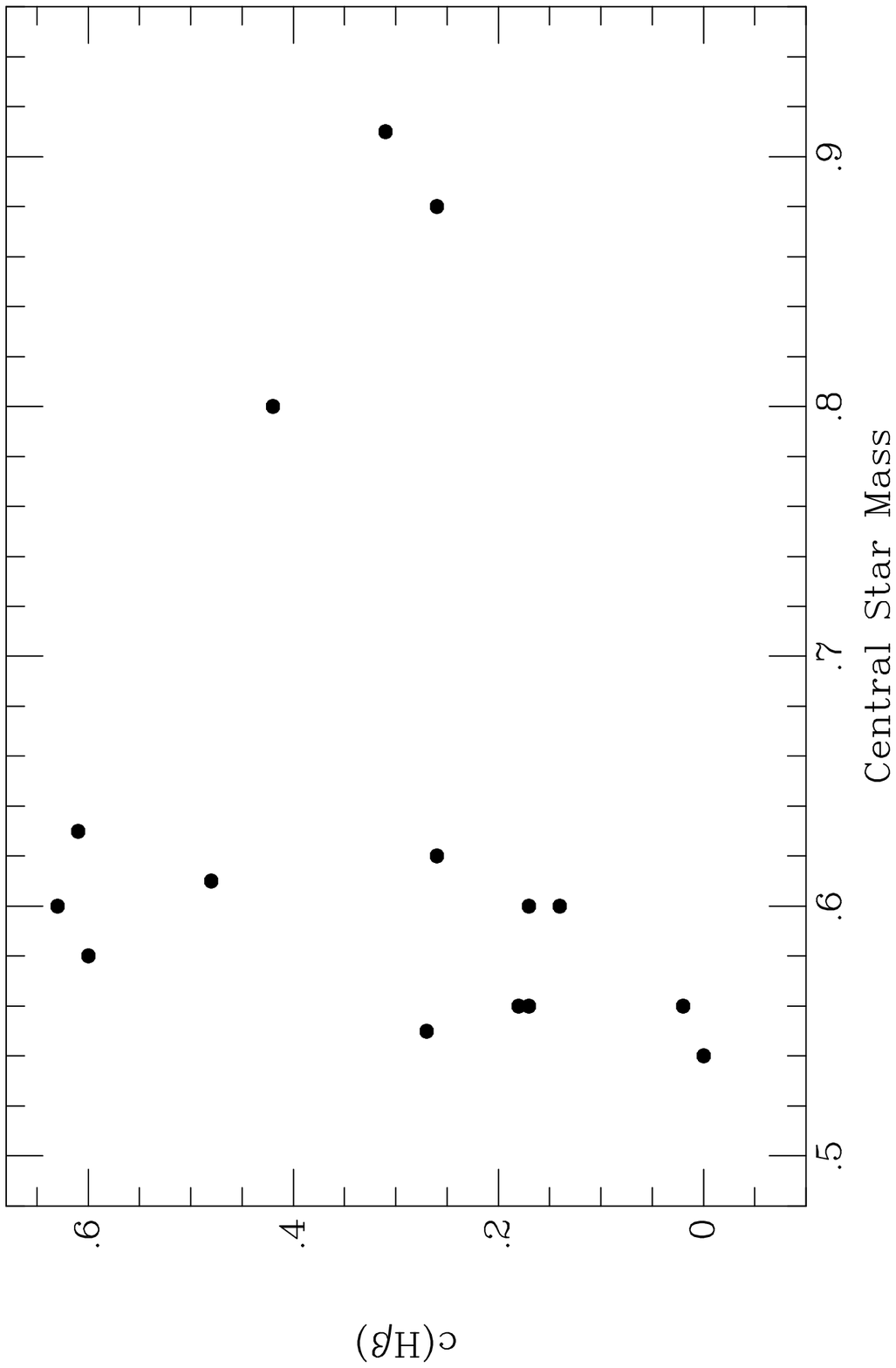}{5.0in}{-90.}{80.}{80.}{-330}{400}
\caption
{A correlation between extinction and central star mass must exist in order
to explain the good and poor correlations seen between absolute and apparent
magnitudes in the previous figures. }
\end{figure}

\begin{figure}
\plotfiddle{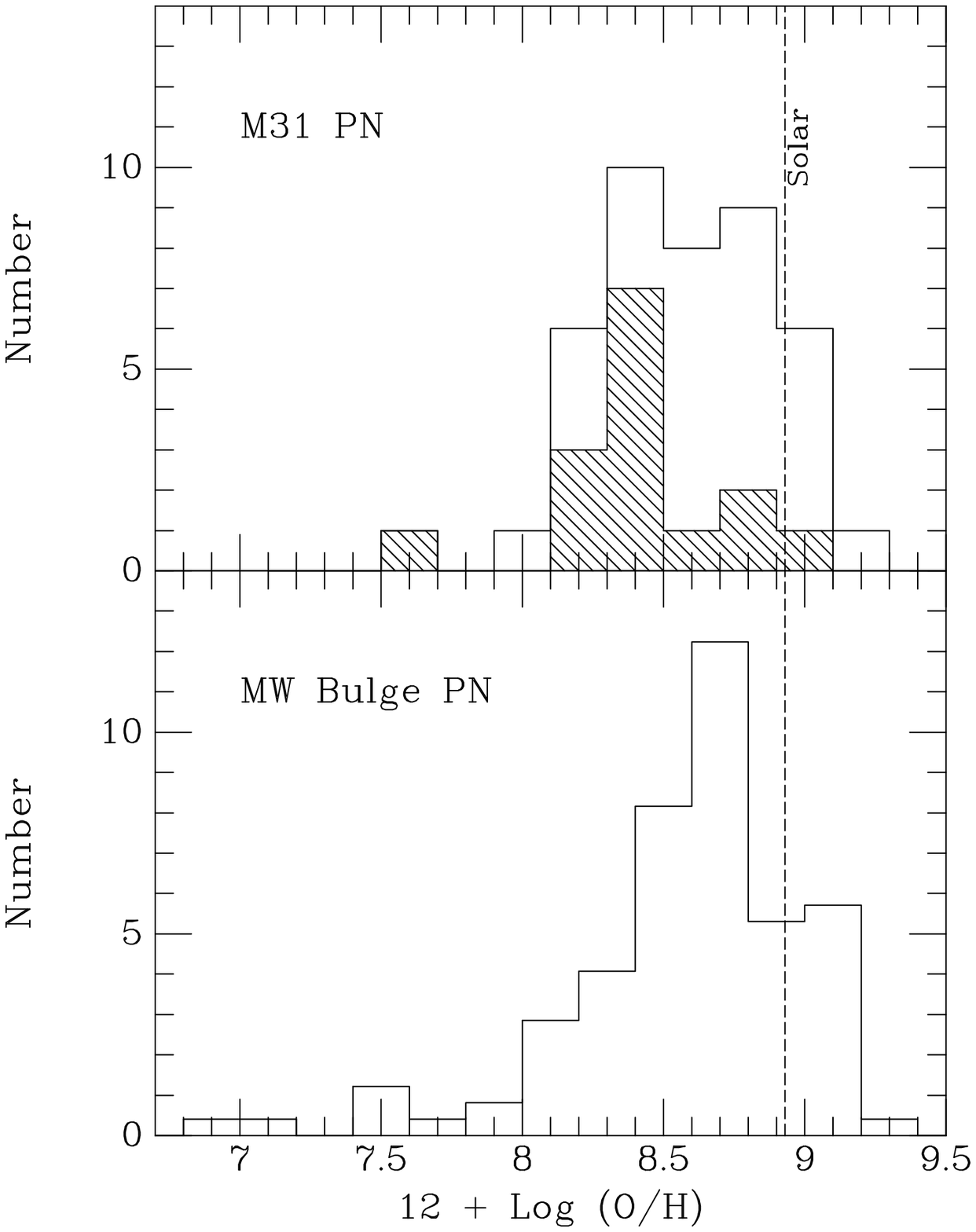}{6.0in}{0.}{80.}{80.}{-240}{-140}
\caption
{The upper panel shows the distribution of oxygen abundances for the 15
PNe in our sample (hatched) and in the total sample (including the SRM
data). The solar oxygen value is indicated by the vertical dashed line.
The SRM sample draws from a somewhat more oxygen-rich population than our
sample, presumably because of that sample's bias toward's [OIII]-bright
PNe. The lower panel shows the oxygen abundance distribution for the
Galactic bulge PNe from Ratag \etal (1997). That sample of 103 PNe
has been normalized to match the same total number (42) of PNe in
the upper sample (ours plus SRM).  Clearly, few PNe in any sample have
super-solar oxygen abundances.}
\end{figure}

\begin{figure}
\plotfiddle{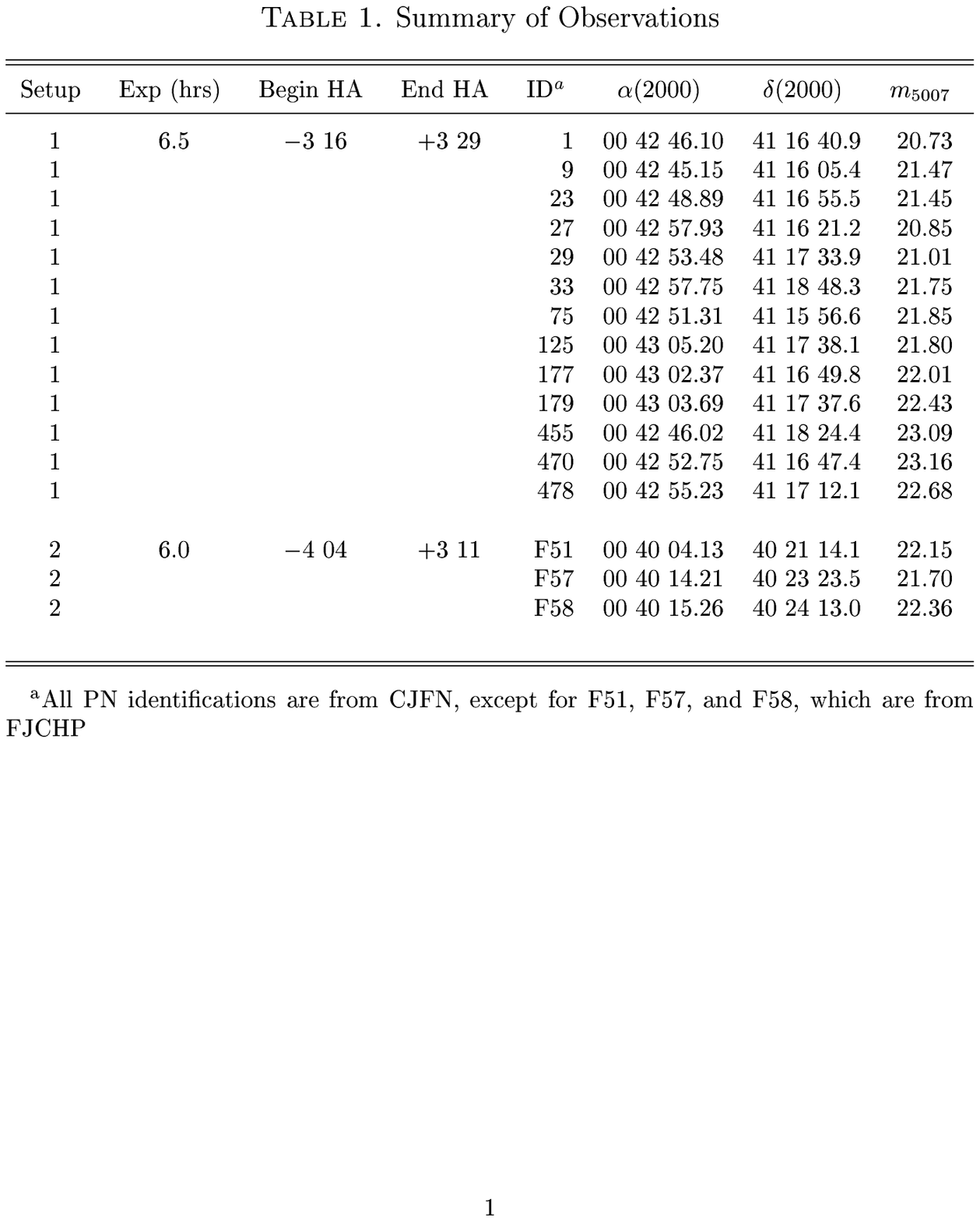}{7.0in}{0.}{100.}{100.}{-300}{0}
\end{figure}

\begin{figure}
\plotfiddle{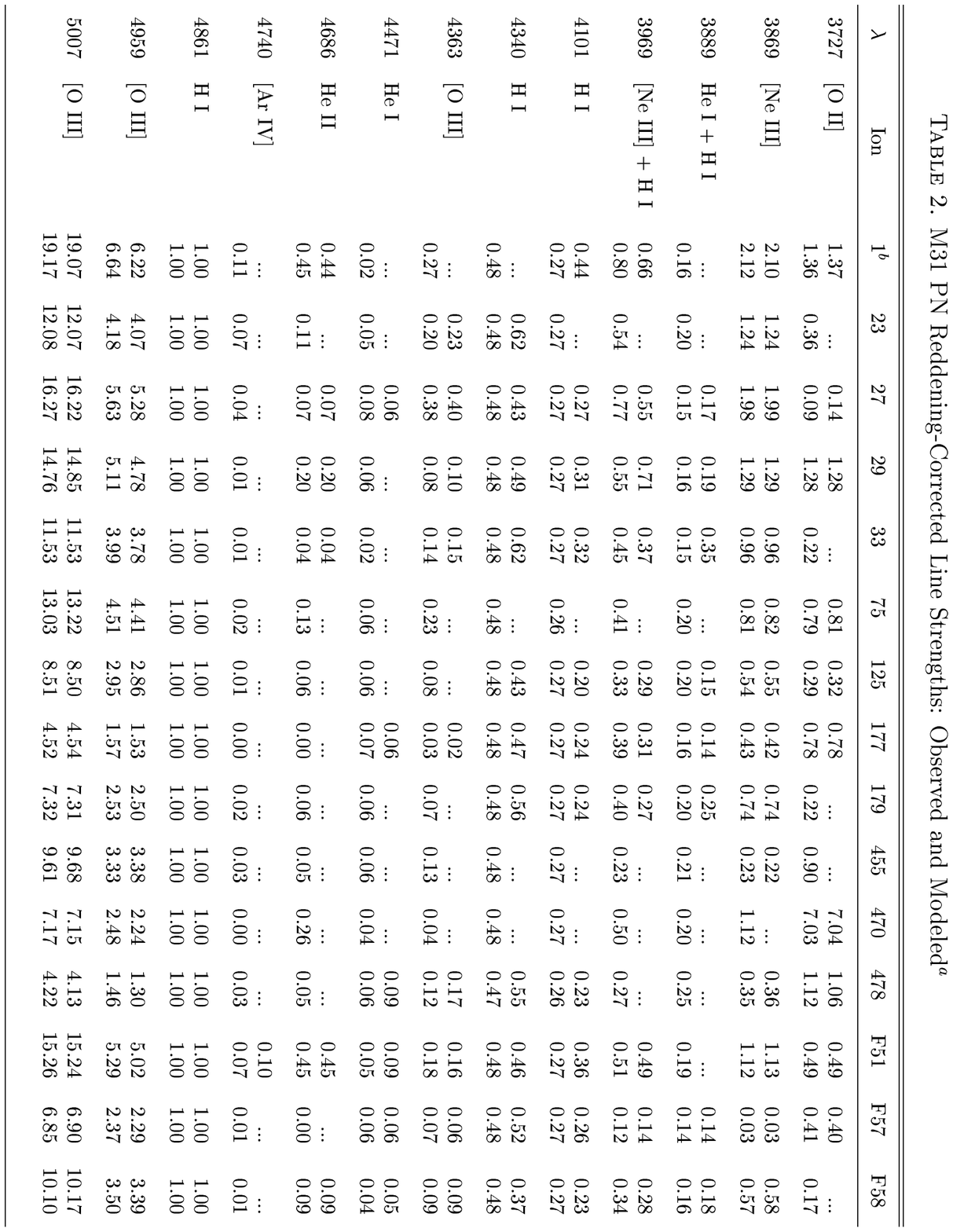}{7.0in}{0.}{100.}{100.}{-320}{-180}
\end{figure}

\begin{figure}
\plotfiddle{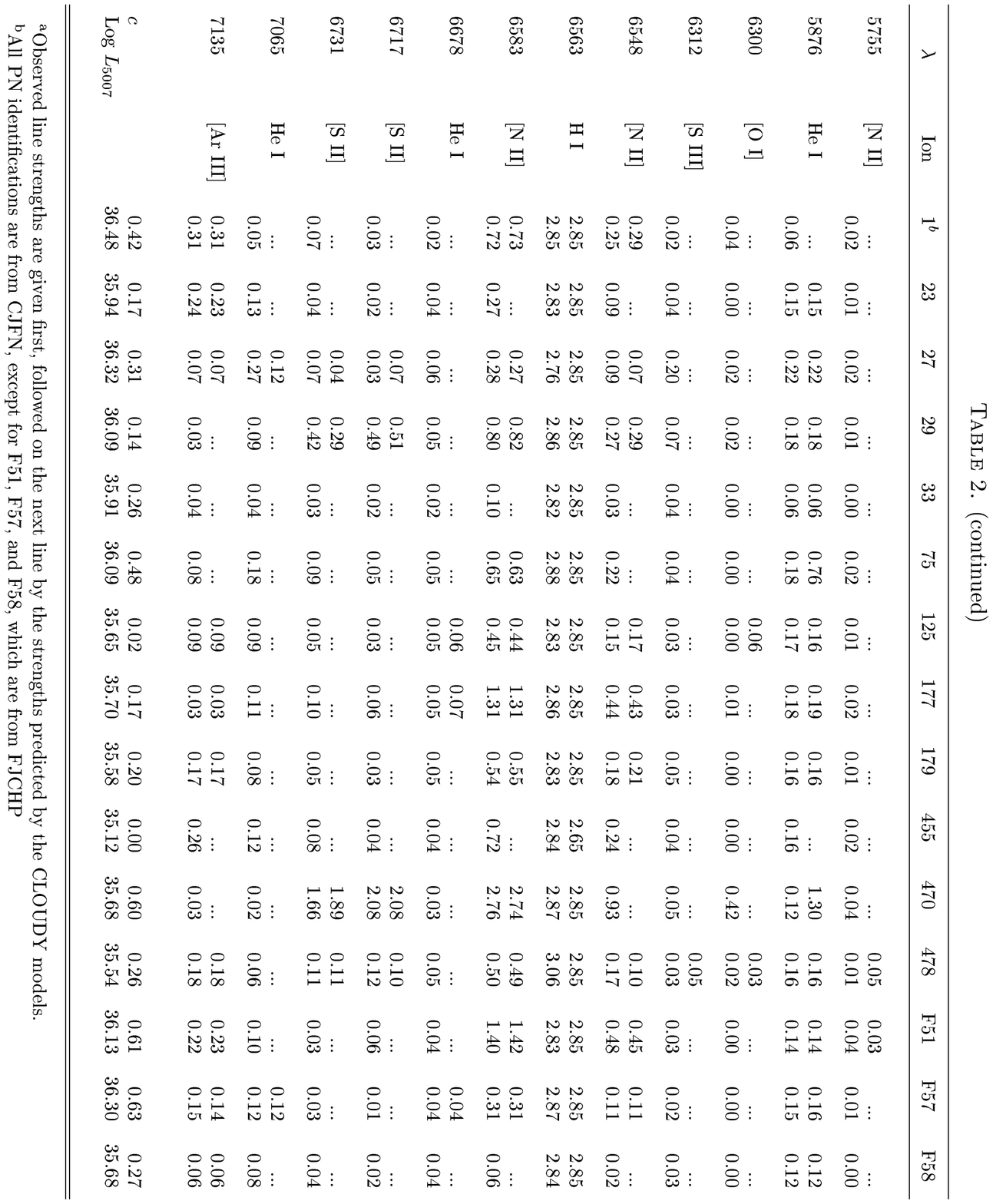}{7.0in}{0.}{100.}{100.}{-320}{-180}
\end{figure}

\begin{figure}
\plotfiddle{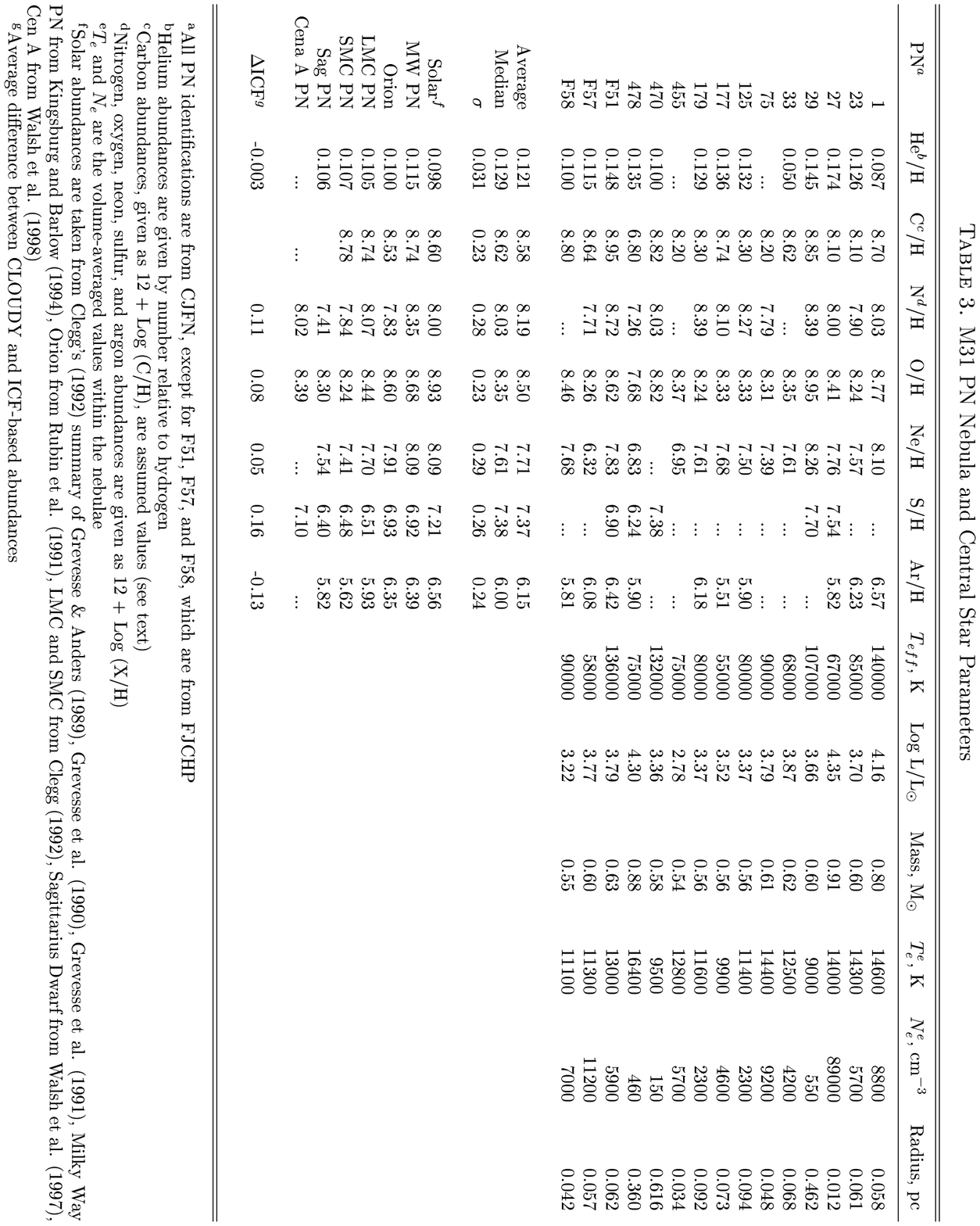}{7.0in}{0.}{100.}{100.}{-320}{-165}
\end{figure}

\end{document}